\documentclass[aps,prb,twocolumn,superscriptaddress]{revtex4-2}

\usepackage{amsmath}
\usepackage{rotating}
\usepackage{graphicx}
\usepackage{hyperref}
\usepackage{epstopdf}
\usepackage{booktabs}
\usepackage{longtable}
\usepackage{float}

\usepackage{colortbl}
\usepackage{natbib}
\usepackage{hyperref}
\usepackage{soul}

\usepackage[dvipsnames]{xcolor}

\begin{document}


\title{Kondo screening in Co adatoms with full Coulomb interaction}


\author{Angelo Valli}
\affiliation{Institut f\"ur Theoretische Physik, Technische Universit\"at Wien, Wiedner Hauptstrasse 8-10, 1040 Vienna, Austria}
\author{Marc Philipp Bahlke}
\affiliation{Institut f\"ur Anorganische und Angewandte Chemie, Universit\"at Hamburg, Martin-Luther-King-Platz 6, 20146 Hamburg, Germany}
\author{Alexander Kowalski}
\affiliation{Institut f\"ur Theoretische Physik und Astrophysik and W\"urzburg-Dresden Cluster of Excellence ct.qmat, Universit\"at W\"urzburg, 97074 W\"urzburg, Germany}
\author{Michael Karolak}
\affiliation{Institut f\"ur Theoretische Physik und Astrophysik, Universit\"at W\"urzburg, Am Hubland, 97074 W\"urzburg, Germany}
\author{Carmen Herrmann}
\affiliation{Institut f\"ur Anorganische und Angewandte Chemie, Universit\"at Hamburg, Martin-Luther-King-Platz 6, 20146 Hamburg, Germany}
\author{Giorgio Sangiovanni}
\affiliation{Institut f\"ur Theoretische Physik und Astrophysik and W\"urzburg-Dresden Cluster of Excellence ct.qmat, Universit\"at W\"urzburg, 97074 W\"urzburg, Germany}
\email[]{Your e-mail address}


\date{\today}

\begin{abstract}
Using a numerically exact first-principles many-body approach, 
we revisit the ``prototypical'' Kondo case of a cobalt impurity on copper. 
Even though this is considered a well understood example of the Kondo effect, 
we reveal an unexpectedly strong dependence of the screening properties 
on the parametrization of the local Coulomb tensor. 
As a consequence, the Kondo temperature can vary by orders of magnitude 
depending on the complexity of the parametrization of the electron-electron interaction. 
Further, we challenge the established picture of a spin-1 moment 
involving two cobalt $d$-orbitals only, 
as orbital-mixing interaction terms boost the contribution of the remainder of the $d$-shell.
\end{abstract}

\pacs{}

\maketitle

\section{Introduction} 

The Kondo effect arises when a local magnetic moment is quantum mechanically
screened by the conduction electrons of a metallic host.
Explained by Jun Kondo in the 1960s~\cite{kondo}, 
this phenomenon has been extensively studied thereafter 
within Anderson's poor man's approach 
and Wilson's renormalization group~\cite{hewson}.
As a direct consequence of the screening of the impurity magnetic moment, 
the spin susceptibility
undergoes a crossover from a Curie-Weiss to a Pauli behaviour upon lowering the temperature.
At the same time, the Abrikosov-Suhl-Kondo resonance~\cite{abrikosov65,suhl65,hewson} 
emerges in the electronic spectral function at the Fermi level.  
Magnetic response functions and electron transport are therefore suitable probes of the Kondo effect.
Despite its well-defined characterization, the signatures of the Kondo effect emerge at energy scales of the order 
of the Kondo temperature $T_K$, which is often of the order of a few Kelvin, making the theoretical description of realistic Kondo systems intrinsically hard. 
Further, the Fermi-liquid properties emerging
below the Kondo temperature $T_K$ are typically reached via smooth crossovers 
rather than with sharp transitions, complicating also the experimental detection.

One case of Kondo effect considered to be simple and relatively well understood 
is that of a Co single impurity on a metallic substrate, such as Cu~\cite{wahl02,wahl04,Neel2007,vitaliPRL101}, Au~\cite{wahl04}, and Ag~\cite{wahl04,Meyer2016}. 
In particular, scanning tunneling spectroscopy (STM) has revealed how 
Co adatoms on Cu hosts
display sharp peaks or Fano-like resonances at zero bias~\cite{wahl02,wahl04,Neel2007,vitaliPRL101,frankPRB92,dangPRB93}, 
which are commonly interpreted as a clear experimental signature of the Kondo screening, 
although the origin of these features has been recently challenged~\cite{bouaziz2003.0174}. 
Experimental estimates of the Kondo scale yield, e.g.,  
$T_K \approx 88$~K and $54$~K for Co on Cu(001) and Cu(111), 
respectively~\cite{wahl02,wahl04}. 
 
However, even in the case of a single impurity, 
for transition metal adatoms 
the theoretical description of the Kondo effect is difficult, 
since the whole $d$-shell is likely to play a role in the screening. 
So far, the theoretical understanding of single Co impurities on Cu~\cite{Huang2008,jaco15,surer11,baru15} stresses the main role played 
by two of the five Co-$d$ orbitals. 
In the case of Co/Cu(001) -- on which we shall focus below -- 
the $d_{xy}$ and $d_{z^2}$ orbitals are \emph{Kondo active}, 
in the sense that they are half-filled and carry a magnetic moment. 
Due to the different symmetry, for Co/Cu(111) 
the $d_{z^2}$ orbital is instead fully occupied, 
and the magnetic moment arises from one of the two doublets 
with $E$ symmetry~\cite{Huang2008,baru15}. 
In general, different crystalline environments determine 
variations in the local electronic structure of the impurity 
and lead to drastically different Kondo resonance line shapes 
observed in STM experiments~\cite{Huang2008,vitaliPRL101}. 
Furthermore, the many-body nature of the Kondo effects 
manifests itself also in a strong dependence of $T_K$ 
on the occupation of the Co $3d$ shell~\cite{wahl04,vitaliPRL101,jaco15,dangPRB93}. 
This also means that the hybridization and the charge transfer 
between the impurity and the substrate play an important role.  
This is reflected in a strong dependence of $T_K$ 
on, e.g., the adatom adsorption distance~\cite{baru15,Bahlke2018}, 
in agreement with the experiments~\cite{Neel2007,vitaliPRL101}. 
In general, the Kondo scale depends exponentially 
on the parameters of the theoretical model, 
making reliable estimates of $T_K$ extremely hard. 

For the same reason, it is also difficult 
to exactly pinpoint the details of the physical processes 
underlying the Kondo screening in these systems. 
Interestingly, theoretical calculations indicate that 
the spin state of Co is $S=1$ 
on both the Cu(001) and Cu(111) surfaces~\cite{baru15}. 
However, evidence for very different Kondo scales 
for the Kondo-active orbitals of Co/Cu(001)~\cite{baru15,jaco15} 
suggest an underscreened 
(or possibly a two-stage~\cite{posazhennikova05}) Kondo effect 
to take place, while a single $T_K$ is expected 
for the magnetic doublet of Co/Cu(111), 
although the degeneracy could be lifted 
by, e.g., spin-orbit coupling~\cite{baru15}. 
On the other hand, Nevidomskyy and Coleman~\cite{NCPRL103} showed that, 
in the case of a multi-orbital impurity, 
the stabilization of an impurity high-spin state 
due to Hund's coupling leads 
to a strong reduction of the Kondo coupling, 
and consequently of $T_K$, with respect to 
the spin $S=1/2$ case~\cite{wanPRB51,izumidaJPSJ67}. 
Robust numerical evidence that the Nevidomskyy-Coleman scenario is realized in idealized model systems comes, e.g., from Ref.~\cite{NRG2005}. 
This seems, however, at odds 
with the relatively high estimates of $T_K$ for these systems 
emerging from experiments~\cite{wahl02,wahl04,Neel2007}. 
Hence, the question is whether or not, or under which conditions, 
the Kondo screening of Co 
on a Cu substrate can be described this way upon cooling. 

We identify two key players which may affect the mechanism 
of the Kondo screening, i.e., multi-orbital correlation effects 
arising from the full treatment of the Co $3d$ shell,  
rather than restricting the description to the Kondo-active orbitals only, 
and the approximation of the form the Coulomb interaction. 
Using a combination of density functional theory (DFT) and numerically exact quantum Monte Carlo (QMC) we analyze the many-body processes leading to the formation and the screening of the local moment 
on a Co impurity on Cu(001) in its full realistic complexity. 
We provide a comparative analysis 
of the role of the parametrization of the Coulomb interaction, 
which is so-far scarcely investigated in a systematic way. 
In particular, we take into account the full Coulomb tensor 
in the whole Co $3d$ multiplet, 
hitherto either simplified~\cite{baru15,jaco15} 
or included only at high temperatures~\cite{surer11,gorelovPRB80,dangPRB93},  
and we push our calculations down to temperatures 
which are relevant to the Kondo screening. 

The paper is organized as follows:
In Sec.~\ref{sec:methodology} we provide the details of the ab-initio 
and many-body calculations for Co/Cu(001). 
In Sec.~\ref{sec:scenarios} we discuss the possible Kondo scenarios, 
and in Sec.~\ref{sec:results} we analyze the screening properties, 
providing evidence which supports   
the important role played by the approximations of the Coulomb tensor. 
Finally, Sec.~\ref{sec:discussion} contains a discussion 
of our results in light of previous studies in the literature, 
as well as our conclusions. \\

\begin{figure}[t]
\centering
\includegraphics[width=0.4\textwidth]{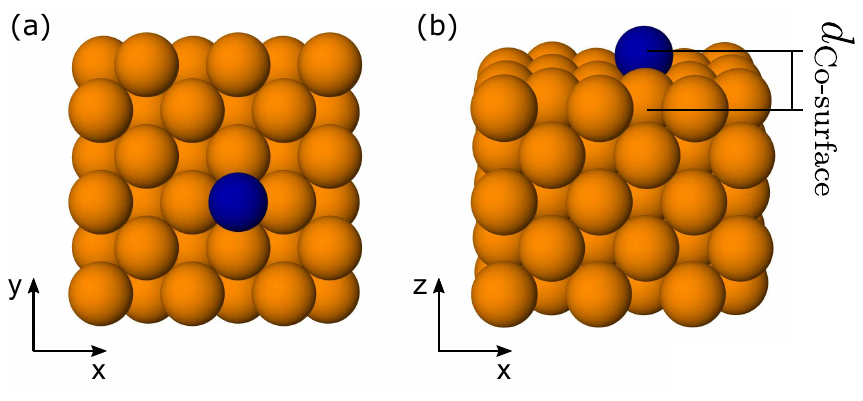}
\caption{Top (a) and side (b) view of the Co/Cu(001) unit cell as used in this study. 
The adsorption distance between the Co adatom and the substrate 
is set to $d_{\textrm{Co-surface}}=1.52$~\AA~\cite{Bahlke2018}.} 
\label{fig:supercell}
\end{figure}

\section{Methodology} \label{sec:methodology}

\subsection{DFT++}\label{sec:DFT++}
Here we investigate the correlation effects 
of a single Co adatom on a Cu(001) surface 
using a combination of DFT within the local density approximation (LDA) 
and the numerical solution of an Anderson impurity model (AIM) 
with realistic parameters. 
This approach is commonly referred to as DFT++ 
in the literature~\cite{Lichtenstein1998,Schueler2017}.

The DFT calculations have been performed with 
the Vienna ab-initio simulation program (VASP)~\cite{Kresse1996,Kresse1999} 
using the projector augmented plane wave (PAW) basis set. 
We modeled the Cu(001) surface as a 4$\times$4 slab consisting of five Cu layers 
using the experimental~\cite{wyckoff} lattice constant of $3.615$~\AA. 
The Co adatom was placed in the fourfold-hollow position  
(see Fig.~\ref{fig:supercell}) 
at an adsorption distance $d_{\textrm{Co-surface}}=1.52$~\AA \ 
with respect to the first Cu(001) layer, 
which we identified in one of our earlier works~\cite{Bahlke2018} 
to be the energetically-favored distance, in agreement with previous literature~\cite{Huang2008}. 
We used a $k$-mesh centered around the $\Gamma$-point 
of size $100 \times 100 \times 1$ $k$-points 
in order to achieve a sufficiently accurate description 
of our Cu(001) substrate. 
This will be necessary for the parametrization of the AIM, 
especially at low temperatures (this important aspect 
is discussed in Appendix~\ref{app:kmesh} in more detail).

\begin{figure}[b]
\centering
\includegraphics[width=0.45\textwidth]{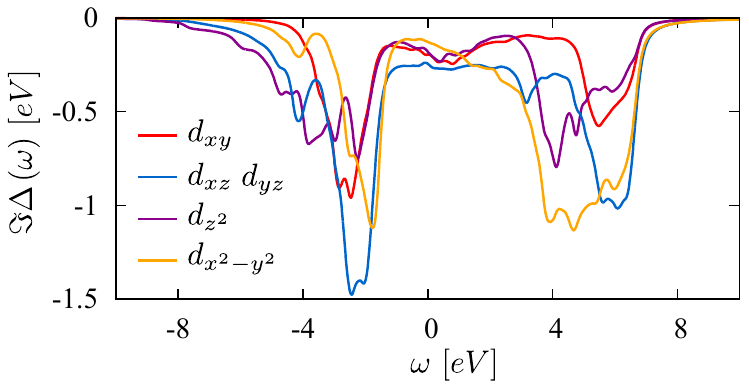}
\caption{Orbital-resolved hybridization function 
$\Im \Delta(\omega)$ for Co/Cu(001). 
The corresponding values at the Fermi level, $\Gamma = -\Im\Delta(0)$ for each orbital 
are reported in Table~\ref{tab:AIM}.}
\label{fig:hyb}
\end{figure}

With the combination of DFT and an AIM, we can take into account 
the correlation effects on the Co atom explicitly 
as well as the realistic complexity 
of its hybridization with the Cu substrate. 
The Hamiltonian of the AIM reads 
\begin{widetext}
\begin{equation}\label{eq:SIAM}
\hat{H} = 
   \sum_{\nu\sigma} \epsilon_{\nu} 
       \hat{c}^{\dag}_{\nu\sigma} \hat{c}^{\phantom{\dag}}_{\nu\sigma} 
 + \sum_{\nu i\sigma} 
   \big( V_{\nu i} \hat{c}^{\dag}_{\nu\sigma} \hat{d}^{\phantom{\dag}}_{i\sigma}
       + V^*_{\nu i} \hat{d}^{\dag}_{i\sigma} \hat{c}^{\phantom{\dag}}_{\nu\sigma} 
   \big) 
 + \sum_{i\sigma} \epsilon_{i} \hat{d}^{\dag}_{i\sigma}\hat{d}^{\phantom{\dag}}_{i\sigma} 
 + \frac{1}{2} \sum_{ijkl} \sum_{\sigma\sigma'}
         U_{ijkl} \hat{d}^{\dag}_{i\sigma} \hat{d}^{\dag}_{j\sigma'}
                  \hat{d}^{\phantom{\dag}}_{l\sigma'}\hat{d}^{\phantom{\dag}}_{k\sigma},
\end{equation}
\end{widetext}
where $\hat{c}^{\dag}_{\nu\sigma}$ ($\hat{c}^{\phantom{\dag}}_{\nu\sigma}$) denotes 
the creation (annihilation) operators for an electron 
with spin $\sigma$ in the $\nu$th bath state 
(in this work, the Cu surface) with energy $\epsilon_{\nu}$, whereas 
$\hat{d}^{\dag}_{i\sigma}$ ($\hat{d}_{i\sigma}^{\phantom{\dag}}$) 
denotes the corresponding operators 
for the $i$th localized $3d$ orbital of the impurity  
(in this work, the Co $3d$ shell) with energy $\epsilon_{i}$. 
The bath and impurity electrons are coupled 
via the hybridization $V_{\nu i}$. 
For QMC techniques, it is convenient reformulate 
the AIM~(\ref{eq:SIAM}) in the action formalism, 
and integrate out the degrees of freedom of the bath 
to obtain a retarded hybridization function 
\begin{equation}
 \Delta_{i}(\omega) =  \sum_{\nu} \frac{V_{\nu i}V^*_{\nu i}}{\omega + \imath 0^+ -\epsilon_i}
\end{equation}
which effectively embeds the impurity into the substrate. Our results, shown in Fig~\ref{fig:hyb} 
are compatible with others found in the literature~\cite{jaco15}.
The hybridization function is then transformed into the Matsubara representation 
$\Delta_{i}(\omega) \rightarrow \Delta_{i}(\imath\omega_n)$ for QMC sampling. 
The values of the orbital-dependent effective crystal fields 
$\epsilon_{i}+ \Re\Delta_i(\infty)$
and hybridization to the substrate $\Gamma_i = -\Im\Delta_i(0)$ 
are given in Table~\ref{tab:AIM} for reference. 

Finally, the tensor 
\begin{equation}
 U_{ijkl} = \int drdr' \psi_i^*(r)\psi_j^*(r') 
                       \frac{e^2}{|r-r'|}
                       \psi_k(r)\psi_l(r')
\end{equation}
describes the local Coulomb interaction 
(we dropped the spin indices, for simplicity) 
as introduced by Slater~\cite{slater1960}, 
with $\psi_{\alpha}$ ($\alpha = i,j,k,l$) being in general any atom-centered basis function 
and $e^2\frac{1}{|r-r'|}$ the long-range Coulomb potential.

\begin{table}[t]
\caption{Single-particle DFT parameters defining the AIM for Co/Cu(001), i.e., 
the effective crystal field $\epsilon_i + \Re\Delta_i(\infty)$ 
and the coupling to the substrate  $\Gamma_i = -\Im\Delta_i(0)$ 
for each orbital in the Co~$3d$ shell. 
The $C_{4v}$ point-group symmetry is enforced at the DFT level, 
so that $d_{xz}$ and $d_{yz}$ are degenerate.
In addition to the crystal field, we also include a double counting correction 
$\mu_{\text DC}=28.0$~eV to constrain the occupation of the $3d$ shell to $n_d=8.0$ electrons. 
}
\label{tab:AIM}
\begin{ruledtabular}
\begin{tabular}{cccc}
Co~$3d$ orbital & $\epsilon_i + \Re\Delta_i(\infty)$~[eV] & $\Gamma_i$~[eV] & \\ 
\colrule
$d_{xy}$          & -0.226 & 0.196 & \\  
$d_{xz}$          & -0.403 & 0.244 & \\  
$d_{z^2}$        & -0.295 & 0.180 & \\  
$d_{yz}$          & -0.403 & 0.244 & \\  
$d_{x^2-y^2}$ & -0.221 & 0.128 & \\   
\end{tabular}
\end{ruledtabular}
\end{table}

\subsection{Coulomb tensor}\label{sec:Coulomb}
The last term of Eq.~(\ref{eq:SIAM}) describes 
the local interaction within the impurity $3d$ shell. 
The full Coulomb interaction $U_{ijkl}$ is in general 
a four-index tensor, which, in the language of second quantization, 
corresponds to different combinations of the 
creation and annihilation operators of the two-body interaction. 
However, due to the extreme numerical complexity required 
to take into account all possible four-fermion terms, 
it is common practice so far to consider approximate interaction schemes. 
Therefore, most previously published results have been obtained 
neglecting -- in a non-systematic and uncontrolled way -- 
parts of the Coulomb tensor. 
With the advent of continuous-time quantum Monte Carlo (CT-QMC) 
methods (see Ref.~\cite{Gull2011} for a review), 
it has become possible to treat the full Coulomb interaction 
without approximations. 
There are already indications in the literature~\cite{gorelovPRB80}  
that the structure of the full Coulomb interaction 
is important to describe the physics of Co/Cu(001). 
We will show that different parametrizations of the Coulomb interaction 
also give rise to substantially dissimilar Kondo screening properties. 

Below we describe the properties of $U_{ijkl}$ in different approximation schemes. 
Within the simplest parametrization, 
one retains only the ``density-density'' terms, 
i.e., those in which the four operators are contracted 
in pairs of number operators 
$\hat{n}_{i\sigma} = \hat{d}^{\dag}_{i\sigma}\hat{d}^{\phantom{}}_{i\sigma}$. 
Within the density-density approximation, 
the Coulomb tensor in Eq.~(\ref{eq:SIAM}) reduces to 
\begin{equation}\label{eq:density-density}
\hat{H}_{\mathrm{D}} = 
   \sum_{i} 
         U_{ii}
         \hat{n}_{i\uparrow}\hat{n}_{i\downarrow} 
 + \sum_{i \neq j} \sum_{\sigma\sigma'} 
        (U_{ij}-J_{ij}\delta_{\sigma\sigma'}) 
         \hat{n}_{i\sigma} \hat{n}_{j\sigma'} .
\end{equation}
In terms of the Coulomb tensor, the above parameters 
$U_{ii}=U_{iiii}$ and $U_{ij}=U_{ijij}$, 
denote the intra- and inter-orbital (direct) interactions, 
while $J_{ij} = U_{ijji}$ denotes 
the density-density Hund's exchange coupling for $\sigma=\sigma'$
(see Supplementary Material for all definitions and symmetry relations). \\
Including also the missing two-body scattering terms,  
which describe ``spin-flip'' ($J_{ij}=U_{ijji}$ for $\sigma\neq\sigma')$ 
and ``pair-hopping'' ($J_{ij}=U_{iijj}$ for $\sigma\neq\sigma'$) processes  
between electrons on different orbitals, 
gives rise to the so-called ``Kanamori'' parametrization, 
of the form  
\begin{equation}\label{eq:kanamori}
\hat{H}_{\mathrm{K}} = \hat{H}_{\mathrm{D}}
                     + \sum_{i\neq j} J_{ij}
        \big(
             \hat{d}^{\dag}_{i\uparrow}
             \hat{d}^{\dag}_{j\downarrow}
             \hat{d}^{\phantom{\dag}}_{i\downarrow}
             \hat{d}^{\phantom{\dag}}_{j\uparrow} 
           - \hat{d}^{\dag}_{i\uparrow}
             \hat{d}^{\dag}_{i\downarrow}
             \hat{d}^{\phantom{\dag}}_{j\uparrow}
             \hat{d}^{\phantom{\dag}}_{j\downarrow}
        \big) ,
\end{equation}
which has the important consequence of restoring 
the rotational invariance of the Coulomb interaction. 
Finally, the ``full Coulomb'' interaction, given by the generic form
\begin{equation}\label{eq:coulomb}
\hat{H}_{\mathrm{C}} = \frac{1}{2} \sum_{ijkl} \sum_{\sigma\sigma'} 
                         U_{ijkl} 
                        \hat{d}^{\dag}_{i\sigma} 
                        \hat{d}^{\dag}_{j\sigma'}
                        \hat{d}^{\phantom{\dag}}_{l\sigma'}
                        \hat{d}^{\phantom{\dag}}_{k\sigma} ,
\end{equation}
contains all possible terms allowed on the $3d$ shell, 
without restrictions. 
In the case of a spherically symmetric atom, 
these terms can be described 
in terms of the Slater radial integrals~\cite{slater1960, Slater1929} 
$F^0$, $F^2$, and $F^4$.
With a spherically-symmetric Coulomb tensor, 
one has the advantage of excluding sources of differences 
associated with specificities of the Cu(001) substrate, 
at the same time allowing us to reduce 
the number of interaction parameters to two: 
$U=F^0$ and $J=\frac{1}{14}(F^2+F^4)$. 
For instance, the intra-orbital Hubbard repulsion  
becomes independent of the orbital index $i$ 
and is given by the relation 
$U_{ii} = F^0+\frac{8}{7}\frac{1}{14}(F^2+F^4)$. 
The different angular dependence of the five $d$-orbitals results 
in four different Hund's couplings $J_{ij}$, 
which can all be expressed in terms of $F^2$ and $F^4$,  
so that $U_{ij}=(U_{ii}+U_{jj})/2-2J_{ij}$ 
(see, e.g. Refs.~\cite{hausoelNC8,karolakPhD} 
and Appendix~\ref{app:tensor} for a detailed discussion). 

\subsection{Details of the Co/Cu(001) calculations}
We solve the AIM~(\ref{eq:SIAM}) by using 
the numerically exact CT-QMC method as implemented 
in the \textsc{w2dynamics} package~\cite{Parragh2012,Wallerberger2018}.
With the choices of interaction parameters $U=4.0$~eV, $J=0.9$~eV,  
and ratio $F^4/F^2 = 0.625$, 
which completely determine the Coulomb tensor, 
the values we use in this study are very similar 
(although spherically symmetric) to those calculated 
for Co/Cu(001) from first principles by Jacob 
within the constrained random phase approximation~\cite{jaco15}. 
For the purpose of showing how significant the differences 
between the results obtained within various interaction schemes can be, 
we compare the magnetic properties of Co/Cu(001) obtained 
by solving the impurity problem with the 
Coulomb interaction $\hat{H}_{\mathrm{C}}$ of Eq.~(\ref{eq:coulomb}), 
as well as with its density-density and Kanamori approximations 
of Eqs.~(\ref{eq:density-density}) and (\ref{eq:kanamori}), respectively. 
We will show that different approximations of the Coulomb tensor 
lead to different physical pictures. 
In particular, the lowest temperature reached here for Co/Cu(001) 
in the scope of the full Coulomb interaction is $T \simeq 33$~K, 
which is below the experimental estimates of $T_K$ for this system.  
To the best of our knowledge, it is the first time that the analysis 
of a single Co adatom on Cu has been pushed to such low temperature 
in the framework of a five-orbital AIM with the full Coulomb interaction.

\begin{figure*}[th]
\centering
\includegraphics[width=\textwidth]{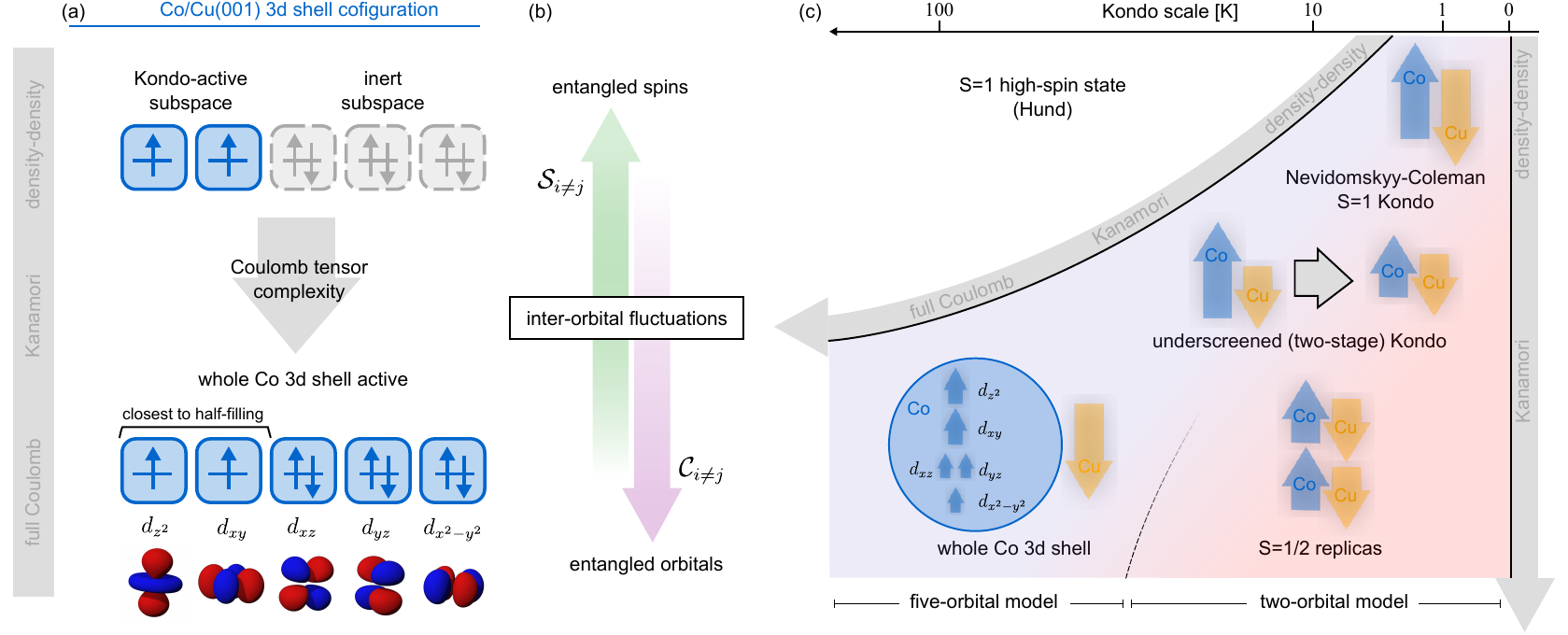}
\caption{
Schematics of possible Kondo scenarios for Co impurities on Cu hosts. 
In all three panels (a-c), the complexity of the Coulomb tensor increases 
from top to bottom. 
(a) Configuration of the Co $3d$ shell hybridized to the Cu(001) surface  
in the presence of Coulomb repulsion. For simplified interactions, 
the $(d_{xy}, d_{z^2})$ half-filled orbitals identify the Kondo-active subspace, 
but for realistic Coulomb tensors the whole multiplet becomes relevant to the Kondo screening. 
(b) Behavior of the spin (${\cal S}_{i \neq j}$) and charge (${\cal C}_{i \neq j}$) 
inter-orbital fluctuations in relation to the complexity of the Coulomb tensor. 
Within the density-density approximation, the physics is dominated by the Hund exchange,  
while exchange interactions (e.g., of the form $U_{ijjk}$ or $U_{ijkl}$) 
favour charge fluctuations and weaken the tendency towards a high-spin state. 
(c) Kondo screening processes (top to bottom). 
If the Co impurity is locked in a high-spin state due to the Hund's coupling $J_H=U_{ijji}$ 
the Nevidomskyy-Coleman scenario~\cite{NCPRL103} can be realized at a very low $T_K$. 
As charge fluctuations are enhanced, 
an underscreened (or possibly two-stage) Kondo effect may take place at a higher $T_K$. 
The extreme limit for two-orbital models is a pair of $S=1/2$ Kondo replicas, 
while in a five-orbital model, the charge redistribution within the whole Co $3d$ shell 
can result in a more complicated Kondo effect and in an enhancement of the Kondo scale. } 
\label{fig:scenario}
\end{figure*}

When interfacing many-body and \textit{ab-initio} calculations, 
as within the DFT++ scheme, 
one should also be aware of the so-called double-counting problem, 
which one encounters because part of the correlation energy 
(in this case on the Co $3d$ shell) 
is already taken into account within DFT. 
Usually, one approximates the double-counting value 
from the fully localized limit (FLL)~\cite{sawatzky94}
or the around mean-field (AMF)~\cite{Anisimov1991} methods. 
Here we follow an alternative procedure, and choose the double counting 
in order to fix the Co $3d$ occupation to 
$n_d = \sum_{i\sigma} n_{i\sigma} = 8$ electrons, instead. 
One reason behind this choice is that the system has been investigated 
in several theoretical studies in an STM-like setup~\cite{jaco15,baru15,Choi2017}, 
where it is assumed that Co on Cu(001) has an $S\!=\!1$ spin state  
with an overall Co $3d$ occupation of roughly $n_d\!=\!8$ electrons. 
This was also confirmed by correlated wave-function-based calculations 
where a Co/Cu$_n$ cluster is embedded in a periodic potential~\cite{Huang2008}.
Under the effect of the substrate crystal field, 
the Co $d_{x^2-y^2}$ and the $(d_{xz}, d_{yz})$ doublet are completely full 
while the $d_{xy}$ and $d_{z^2}$ orbitals are both half-filled. 
In this situation, $S\!=\!1$ high-spin configurations 
are expected to be locally dominant, which calls 
for a systematic analysis of the role of the Hund's coupling 
within the different approximations of the Coulomb tensor. 
However, we will also discuss deviations 
from integer filling of the Co $3d$ shell, 
as they are expected to influence 
the screening properties of Co/Cu(001)~\cite{jaco15}.

\section{Possible Kondo scenarios}\label{sec:scenarios}
The goal of this section is to determine how the Kondo screening 
mechanism can be influenced by the parametrization 
of the local Coulomb repulsion on the Co impurity. 
To this end, we are going to analyze in particular the 
finite-temperature spin and charge response functions, 
calculated at the Co site. 
We compare the three interaction schemes discussed in Sec.~\ref{sec:Coulomb} 
(i.e., density-density, Kanamori, and full Coulomb), 
especially focusing on the Co $d_{xy}$ and $d_{z^2}$ orbitals, 
which are identified 
as the Kondo-active orbitals in the literature~\cite{jaco15,baru15}
(note the different orientation of the $xy$ plane 
here compared with these works). 
However, we will claim that more realistic descriptions 
of the Coulomb tensor favor a scenario 
in which also the other $3d$ orbitals play an important role 
in the screening of the Co local moment.

The scheme presented in Fig.~\ref{fig:scenario} 
anticipates the main results of the present paper. 
The electronic configuration of the Co $3d$ shell 
hybridized with the Cu(001) surface is shown in Fig.~\ref{fig:scenario}(a). 
In the Co adatom with $n_d=8.0$ electrons in the $3d$ shell, a high-spin state 
is always realized for temperatures above the Kondo regime. 
We find a link between the form of the Coulomb interaction, 
which strongly affects spin- and charge fluctuations, 
as represented schematically in Fig.~\ref{fig:scenario}(b), 
and the possible mechanism behind the screening of the Co spin, 
indicated in Fig.~\ref{fig:scenario}(c). 
In the simplest approximation scheme, 
i.e., the one of Eq.~(\ref{eq:density-density}), 
in which only the density-density part of the local Coulomb interaction 
is taken into account, the $(d_{xz}, d_{yz}, d_{x^2-y^2})$ subspace 
is almost completely filled, and can be considered inert. 
Due to the strong Hund's coupling within the Kondo-active subspace, 
the Co impurity is locked into an $S=1$ state down to a few K, 
when it can eventually be screened by the conduction electrons of the Cu surface, 
thus realizing the Nevidomskyy-Coleman scenario 
of the suppression of $T_K$ for an $S=1$ Kondo impurity. 
As we increase the complexity of the Coulomb tensor, by including interaction terms 
beyond the density-density approximation 
in the Kanamori parametrization of Eq.~(\ref{eq:kanamori}) 
and in the full Coulomb parametrizations of Eq.~(\ref{eq:coulomb}), 
the most important effect that we observe 
is a progressive breakdown of the $(d_{xy}, d_{z^2})$ $S=1$ high-spin state. 
We can rationalize this effect in terms of two key players: 
i) the enhancement of charge fluctuations within the whole Co $3d$ multiplet, and
ii) the frustration of the spin correlations due to the competition between 
all generalized exchange interactions in the Coulomb tensor, 
e.g., of the form $U_{ijjk}$
This includes the Hund's coupling and the spin-flip processes ($i=k$)  
as well as additional processes beyond the density-density approximation ($i \neq k$), 
or of the form $U_{ijkl}$, with four different orbital indices. 
A thorough discussion of these terms is provided in Appendix~\ref{app:tensor}. 

It is interesting to speculate on the suitable screening mechanisms 
which could replace the Nevidomskyy-Coleman scenario 
for Co/Cu systems, 
in order to look for their characteristics in our numerical analysis. 
One possibility is the underscreened Kondo effect, 
where the Co spin is only partially screened by the substrate. 
Depending on how many modes of the host effectively couple to the  impurity, 
a Nozi\'eres Fermi liquid can be recovered at lower $T$ 
by screening the remaining spin 
(thus realizing a two-stage Kondo effect). 
In the regime where the charge fluctuations become dominant, 
the $d_{xy}$ and $d_{z^2}$ orbitals may also behave 
as a pair of $S=1/2$ replicas, 
which are screened at possibly very different $T_K$s. 
Moreover, depending on the degree of orbital degeneracy 
of the $3d$ multiplet, an SU(4) Kondo effect could also take place. 
The latter may be relevant for the Co/Cu(111) case, 
where the Co magnetic state is actually a doublet~\cite{baru15}.
The increased symmetry, from an SU(2) spin-Kondo 
to an SU(4) spin-orbital Kondo -- or even an SU(N) symmetry, 
involving also the rest of the $3d$ multiplet -- 
is generally expected to result  
in a single enhanced Kondo scale~\cite{sasakiNat405}. 
All the above mechanisms would be compatible 
with the relatively high $T_K \sim 50-100$~K estimated 
by transport experiments~\cite{wahl02,wahl04,Neel2007}. 

While the general role of the Coulomb interaction 
emerges clearly from our calculations, 
a precise estimate of $T_K$ and the identification 
of the Kondo mechanism responsible for the screening 
for each parametrization of the Coulomb tensor remains elusive. 
This is mostly due to the difficulty of observing 
typical Fermi liquid temperature scaling within our methodology.

\section{Results}\label{sec:results}

\subsection{Spin correlations and effective local moment}
\label{sec:spin-spin}
In order to investigate the screening of the impurity magnetic moment 
we sample the spin-spin response function in imaginary time within CT-QMC:
\begin{equation}\label{eq:spin-spin-cor}
\chi_{ij}(\tau) = g^2\langle \hat{S}^z_i(\tau) \hat{S}^z_j(0)\rangle, 
\end{equation}
where $i$ and $j$ denote the Co $3d$ impurity orbitals, 
$\hat{S}^z_i$ is the local spin operator on orbital $i$,  
and $g$ is the electron spin gyromagnetic factor. 
The static (i.e., $\omega=0$) spin susceptibility 
is obtained via integration of the diagonal elements 
of Eq.~(\ref{eq:spin-spin-cor}), as
\begin{equation}\label{eq:static_spin}
 \chi_{ii}(T) = \int_0^\beta d\tau \chi_{ii}(\tau),
\end{equation}
where $\beta$ is the inverse temperature. 
For a Kondo impurity, the static spin susceptibility 
follows a Curie-Weiss behavior $\chi(T) \propto 1/T$  
in the local moment regime well above $T_K$.  
As the moment is screened by the conduction electrons,  
the spin susceptibility has a crossover to a Pauli behavior 
due to the onset of a Fermi liquid (FL) regime: 
$\chi^{-1}(T) \propto T \!+\! T_{FL}$, 
with the characteristic coherence temperature $T_{FL}$  
corresponding to $T_K$ 
in the case of a single impurity~\cite{burdinPRL85,amaricciPRB85}.

\begin{figure}[b]
\centering
\includegraphics[width=0.5\textwidth]{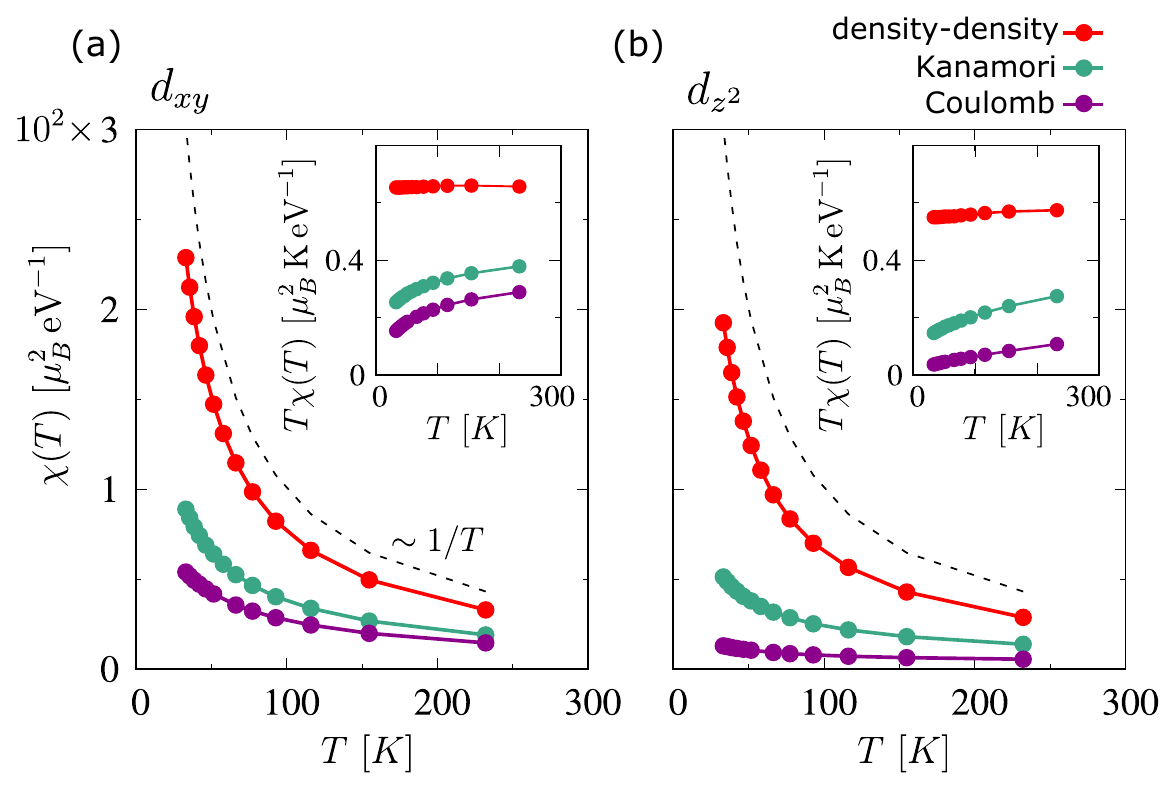}
\caption{Orbital-resolved static spin-spin response function  
$\chi(T) \equiv \chi_{ii}(T)$ 
for the $d_{xy}$ (a) and the $d_{z^2}$ (b) orbitals. 
The dashed lines shows the Curie-Weiss behavior $\chi(T) \propto 1/T$ 
in the local moment regime. 
Plotting $T\chi(T)$ (insets) highlights the differences 
observed with the three interaction schemes, 
with $T\chi(T) \!\sim\! const.$ in the local moment regime, 
and linearly vanishing at $T \ll T_K$ (see text). }
\label{fig:xT_inset}
\end{figure}

\begin{figure}[t]
\centering
\includegraphics[width=0.5\textwidth]{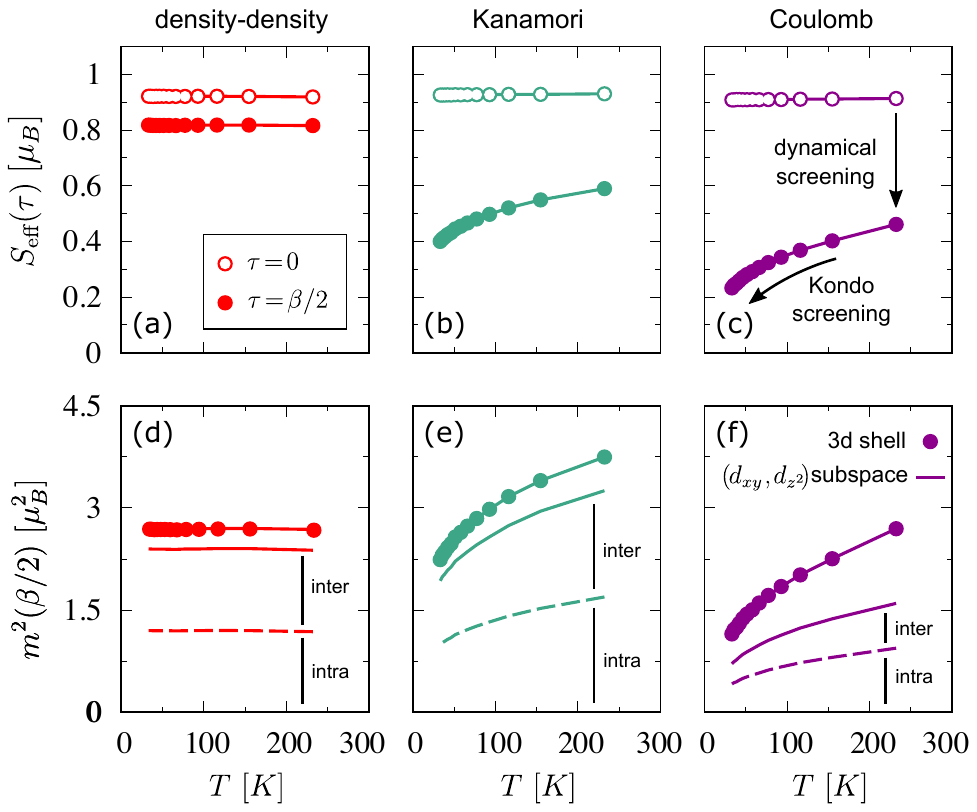} 
\caption{Analysis of the spin correlations at the Co impurity. 
(a-c) Unscreened (open symbols) and screened (filled symbols) 
effective spin $S_{\text{eff}}(\tau)$ 
estimated from the spin susceptibility (see text for the details). 
(d-f) Screened magnetic moment $m^2(\beta/2)$ 
within the whole $3d$ shell (filled symbols) 
and restricted to the $(d_{xy}, d_{z^2})$ subspace (solid line). 
The dashed lines separates intra- and inter-orbital contributions 
within the subspace. 
Due to the increasing contribution of orbitals outside the subspace, 
a two-Kondo-active orbital description of the magnetic moment 
becomes unsatisfactory for realistic Coulomb interactions. }
\label{fig:Seff}
\end{figure}

In Fig.~\ref{fig:xT_inset} we compare the spin susceptibility 
of the $d_{xy}$ and $d_{z^2}$ orbitals 
obtained for all interaction parametrizations. 
In the corresponding inset we also plot $T \chi_{ii}(T)$, 
as it is customarily done in order to represent  
a Curie-Weiss susceptibility as a constant
and a Pauli susceptibility as linearly vanishing. 
This allows to highlight the differences observed 
with the three interaction schemes. 
Within the density-density approximation, we obtain 
an almost perfect $1/T$ behaviour of the susceptibilty 
(in the main panels, and a plateau in the insets) 
for both orbitals, indicating a Curie-Weiss behavior 
in the full range of temperatures of our calculations. 
Consequently, we can infer that the upper bound 
for the Kondo temperature within the density-density approximation 
is substantially lower than $30$~K, 
i.e., it is likely of $\mathcal{O}(1)$~K. 
Instead, the Kanamori and full Coulomb parametrizations 
yield clear signatures of Kondo screening 
in the same temperature window. 
The Kanamori coherence scale seems to be still quite low, 
and at about $30$~K the crossover from a residual entropy 
to a fully screened moment is indeed far from being complete. 
We also observe a pronounced departure from a constant $T\chi(T)$ 
within the Kanamori and ---even more evidently--- the full Coulomb parametrizations. 
A linearly vanishing $T\chi(T)$  is clearly observed for the 
$d_{z^2}$ orbital (yet not for $d_{xy}$ one),  
which suggest different Kondo scales $T_K^{{z^2}}>T_K^{{xy}}$ 
for those two orbitals. 
In order to quantify this analysis, we extract $T_K$ 
from the saturation that characterizes the crossover from Curie-Weiss to Pauli behavior 
of the spin susceptibility. 
We obtain $T_K$ individually for each with the following fitting function~\cite{wilsonRMP47,hausoelNC8}
\begin{equation}
 \chi(T) = \frac{\mu^2}{3k_B(T + 2T_K)},  \nonumber
\end{equation}
where $\mu$ is a fitting parameter of the same order of magnitude of Bohr's magneton $\mu_B$. 
Within the density-density approximation, we estimate $T_K<1$~K for both orbitals. 
A significant enhancement of the Kondo scale is found 
within the Kanamori: 
$T_K^{{xy}}=8.5$~K and $T_K^{{z^2}}=14$~K, 
and within the full Coulomb: $T_K^{{xy}}=18$~K and $T_K^{{z^2}}=40$~K parametrizations.  
In particular, the latter value is also in qualitative agreement with the estimate 
of $T_K \approx 88$~K from the experiments~\cite{wahl02}. 
Recently, it has been shown how a reliable estimate of the Kondo temperature 
can also be obtained from the low-frequency structure of the generalized charge susceptibility, which allows one to identify the hallmarks 
of the formation of the local moment and of the Kondo screening~\cite{chalupa2003.07829}.

Further insight in the different screening processes 
activated by the Coulomb interaction 
can be obtained by inspecting two special values 
of the impurity spin susceptibility in imaginary time: 
$\chi(\tau\!=\!0)$ and $\chi(\tau\!=\!\beta/2)$. 
At $\tau\!=\!0$, it corresponds 
to the (square of the) bare magnetic moment, 
sometimes also called the unscreened paramagnetic moment. 
It indicates the tendency of the Co impurity to build up a quantum magnetic moment at short time scales. 
Instead, its value at $\tau\!=\!\beta/2$ can be associated 
to a magnetic moment at asymptotically long times, 
and hence it provides information on the effectiveness 
of the dynamical screening 
due to quantum fluctuations~\cite{toschiPRB86}. 
These two quantities are helpful to visualize 
the different screening properties 
within the three interaction schemes and allow us to understand 
which two-body processes are decisive for the Kondo screening. 

In a correlated system we expect a strong contribution 
from the orbital off-diagonal components of the spin susceptibility,  
and in particular, in the case under study 
they are equally important as the diagonal ones. 
We inspect the screening properties by looking at the total 
(unscreened and screened respectively) ``effective'' spin moment 
$S_{\textrm{eff}}$. 
This involves all components of $\chi_{ij}(\tau)$ 
and takes into account the difference between the quantum nature 
of the spin degrees of freedom of the three parametrizations 
of the Coulomb interaction. 
Within the density-density approximation we describe an Ising spin, 
so that the (instantaneous, i.e., $\tau \! = \! 0$) magnetic moment is given by 
\begin{equation}\label{eq:spin-Ising}
m^2_{\textrm{Ising}} =g^2 \langle \hat{S}_z^2 \rangle. 
\end{equation}
Instead, since the Kanamori and the full Coulomb parametrizations 
preserve the spin SU(2) rotational invariance of the Coulomb tensor, 
the magnetic moment is given by 
\begin{equation}\label{eq:spin-SU2}
 m^2_{\textrm{Heisenberg}} = g^2 [\langle \hat{S}_x^2 \rangle + 
           \langle \hat{S}_y^2 \rangle + 
           \langle \hat{S}_z^2 \rangle] 
     = 3 g^2 \langle \hat{S}_z^2 \rangle.
\end{equation}
We can hence define 
$m^2\!=\! \xi \sum_{ij} \chi_{ij}(\tau\!=\!0)$, 
where $\xi\!=\!3$, except for the density-density case 
in which $\xi\!=\!1$, and 
the indices $i$ and $j$ in the summation 
run over either all Co $3d$ orbitals, 
or over a subset thereof, as necessary. 
The natural generalization at finite imaginary time is therefore 
\begin{equation}
 m^2(\tau)\!=\!\xi \sum_{ij} \chi_{ij}(\tau), \
 \textrm{with}
 \begin{cases}
    \xi=1,& \textrm{Ising}\\
    \xi=3,& \textrm{Heisenberg}
\end{cases}
\end{equation} 
which allow us to extract the effective spin $S_\text{eff}(\tau)$ 
from the relation $m^2 = g^2 S^2_\text{eff}$ 
for density-density (Ising spin),  
or $m^2 = g^2 S_\text{eff}(S_\text{eff}+1)$ 
for Kanamori and full Coulomb interactions (Heisenberg spin)~\cite{toschiPRB86}.

The empty symbols in the three upper panels of Fig.~\ref{fig:Seff} 
show the unscreened (i.e., instantaneous) 
effective spin $S_{\text{eff}}(\tau = 0)$, 
including the intra- and inter-orbital contributions 
from the whole Co $3d$ shell. 
For all interaction parametrizations we get 
an instantaneous paramagnetic spin moment $S_\text{eff} > 0.9$, 
in excellent agreement with the value $S = 1$ 
expected in the high-spin configuration, 
and with the literature~\cite{jaco15}, 
which remains perfectly constant 
in the whole range of temperatures considered here. 
The screened effective moment $S_{\text{eff}}(\tau = \beta/2)$ 
at each temperature is suppressed with respect to its $\tau=0$ 
counterpart by quantum fluctuations. 
Within the density-density approximation, we observe a sizable effective moment 
$S_\text{eff}(\beta/2) \approx 0.8$ down to the lowest temperature investigated. 
This mirrors the information obtained by the analysis of the static 
spin susceptibility, and substantially rules out 
any temperature-dependent (i.e., Kondo) screening of the local moment 
in this temperature window within the density-density approximation. 
In contrast, within both the Kanamori and the full Coulomb parametrizations 
we observe the pronounced screening of the ``long-time'' local moment, 
which is considerably stronger than what observed within the density-density approximation. 
At the same time, a clear temperature dependence of $S_\text{eff}$ 
indicates a strong ability of the environment 
to Kondo-screen the Co impurity spin. 
Therefore, even at integer filling of the Co $3d$ shell ($n_d=8$), 
the local quantum fluctuations described by more complete 
parametrizations of the Coulomb interaction 
disgregate the high-spin state already in the high-temperature regime, 
and favor the onset of Kondo screening. 
Instead, this does not happen in the density-density case, 
for which the Nevidomskyy-Coleman scenario 
of a strong suppression of $T_K$ for a spin $S=1$ is fully realized. 

We can analyze the orbital character 
of the impurity magnetic moment by looking at the (screened) partial magnetic moment 
(which is an additive quantity, unlike $S_\text{eff}$). 
This is obtained by restricting the double sum over $i$ and $j$ 
in the definition of $m^2$ to the $(d_{xy}, d_{z^2})$ subset of orbitals. 
We can also distinguish between the $m^2_{\textrm{intra}}$ ($i=j$) 
and $m^2_{\textrm{inter}}$ ($i \neq j$) components within the subspace. 
As shown in the lower panels of Fig.~\ref{fig:Seff}, 
within the density-density approximation, $S_{\text{eff}}(\beta/2)$ is mostly determined 
by the $(d_{xy}, d_{z^2})$ subspace, 
whereas in the full Coulomb parametrizations 
there is substantial contribution from the 
$d_{xz}$, $d_{yz}$, and $d_{x^2-y^2}$ orbitals. 
This demonstrates that a two-Kondo-active orbitals description 
of the system is no longer accurate when a realistic Coulomb interaction 
is taken into account. 
The intra- and inter-orbital contributions 
to the local moment within the $(d_{xy}, d_{z^2})$ subspace 
are similar to each other for all three parametrizations, 
but both are strongly suppressed by introducing interaction terms 
beyond the density-density approximation. 
As we discuss in Sec.~\ref{sec:decoupling}, 
this observation can be understood by considering the charge redistribution 
within the whole Co $3d$ multiplet,  
which competes with the spin-locking tendency induced by the Hund's coupling.

\subsection{Spin and charge fluctuations} \label{sec:decoupling}
\begin{figure}[b]
\centering
\includegraphics[width=0.5\textwidth]{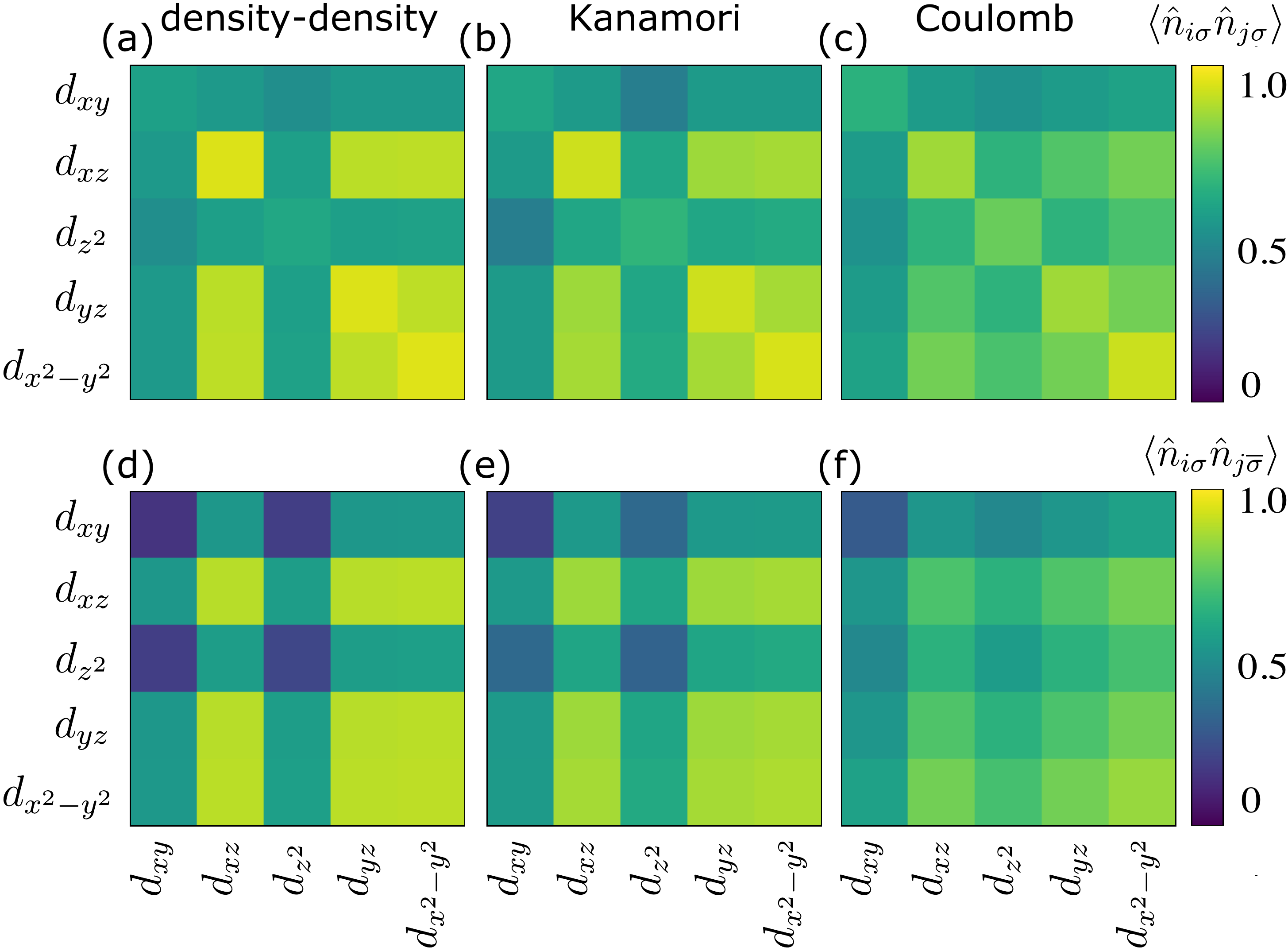}
\caption{Generalized double occupations 
$\langle n_{i\sigma} n_{j\sigma'}\rangle$ 
for parallel ($\sigma'=\sigma$) 
and anti-parallel ($\sigma'=\overline{\sigma}$) spin orientation 
for the Co $3d$ multiplet at $T \approx 33$~K. } 
\label{fig:nisnjs}
\end{figure}

The full Coulomb tensor 
(even if here it still assumes a spherical environment) 
represents the reference point in our comparative analysis, 
as it gives the most coherent of all the results 
and the largest Kondo temperature, meaning the closest to the experiments. 
In order to ascertain the origin of the physical differences between 
the full Coulomb and the two other approximate schemes we consider the generalized double occupations 
$\langle \hat{n}_{i\sigma} \hat{n}_{j\sigma'} \rangle$ 
for parallel ($\sigma'=\sigma$) 
and anti-parallel ($\sigma'=\overline{\sigma}$) spin orientations. 
The numerical data representative of the low-temperature regime 
(at $T \approx 33$~K) are collected in Fig.~\ref{fig:nisnjs} 
and illustrated by a set of matrix heatmaps, 
but there temperature dependence is much weaker 
than their dependence on the parametrization of the Coulomb interaction. 
For $\sigma = \sigma'$, the diagonal elements correspond 
to the spin- and orbital-resolved occupations 
$\langle \hat{n}_{i\sigma} \rangle$. 
Note that all quantities are symmetrized 
over both spin $(\sigma \leftrightarrow \sigma')$ 
and orbital $(i \leftrightarrow j)$ indices. 
Within the density-density approximation, 
both the $d_{xy}$ and $d_{z^2}$ orbitals are close to half-filling 
(i.e., $\langle \hat{n}_{i\sigma} \rangle = 0.5$ electrons) 
and have well defined local moments. 
All the other Co $3d$ orbitals are almost full. 
Moreover, within the $(d_{xy}, d_{z^2})$ subspace, 
$\langle \hat{n}_{i\sigma} \hat{n}_{j\sigma} \rangle \gg 
 \langle \hat{n}_{i\sigma} \hat{n}_{j\overline{\sigma}} \rangle$, 
for $i \neq j$, which marks the clear tendency 
towards a $S=1$ high-spin configuration  
favored by the Hund's coupling $J_H = U_{ijji}$. 
Within this picture, which is very similar to 
the atomic ground state configuration~\cite{Huang2008,baru15}, 
not only can one identify $d_{xy}$ and $d_{z^2}$   
as the Kondo-active orbitals, but one could na\"{i}vely expect 
the physics to be described to a good degree of approximation 
by a two-orbital AIM, as also assumed in previous literature~\cite{baru15}.

The situation is substantially overthrown in the case 
of the Kanamori and full Coulomb parametrizations. 
In fact, by progressively including more interaction terms beyond 
the density-density approximation, i.e., moving from left to right in Fig.~\ref{fig:nisnjs}, 
two trends emerge clearly: 
(i) There is a significant charge redistribution 
within the Co $3d$ shell. In particular $\hat{n}_{i\sigma}$ 
in the $(d_{xy}, d_{z^2})$ subspace increases as 
$(0.57, 0.60) \rightarrow (0.59,0.65) \rightarrow (0.64, 0.77)$,  
resulting in the suppression of the local moment 
of the Kondo-active subspace observed in Fig.~\ref{fig:Seff}. 
(ii) The inter-orbital ($i \neq j$) double occupations 
for parallel and anti-parallel spin orientations 
become progressively more similar, i.e., 
$\langle \hat{n}_{i\sigma} \hat{n}_{j\sigma} \rangle \simeq 
 \langle \hat{n}_{i\sigma} \hat{n}_{j\overline{\sigma}} \rangle$ 
for \emph{all} pairs of orbitals. 
As a consequence, the tendency towards a high-spin state 
of the $(d_{xy}, d_{z^2})$ pair is substantially weakened. 
At the same time, the $d_{xy}$ and $d_{z^2}$ orbitals 
still possess the two largest local moments of the entire multiplet, 
so that they supposedly maintain a prominent role 
in the Kondo screening process, 
but with important contributions to the physics 
coming from the other orbitals. 
The results are in complete agreement with the conclusions 
of the spin susceptibility analysis.

\begin{figure}[b]
\centering
\includegraphics[width=0.5\textwidth]{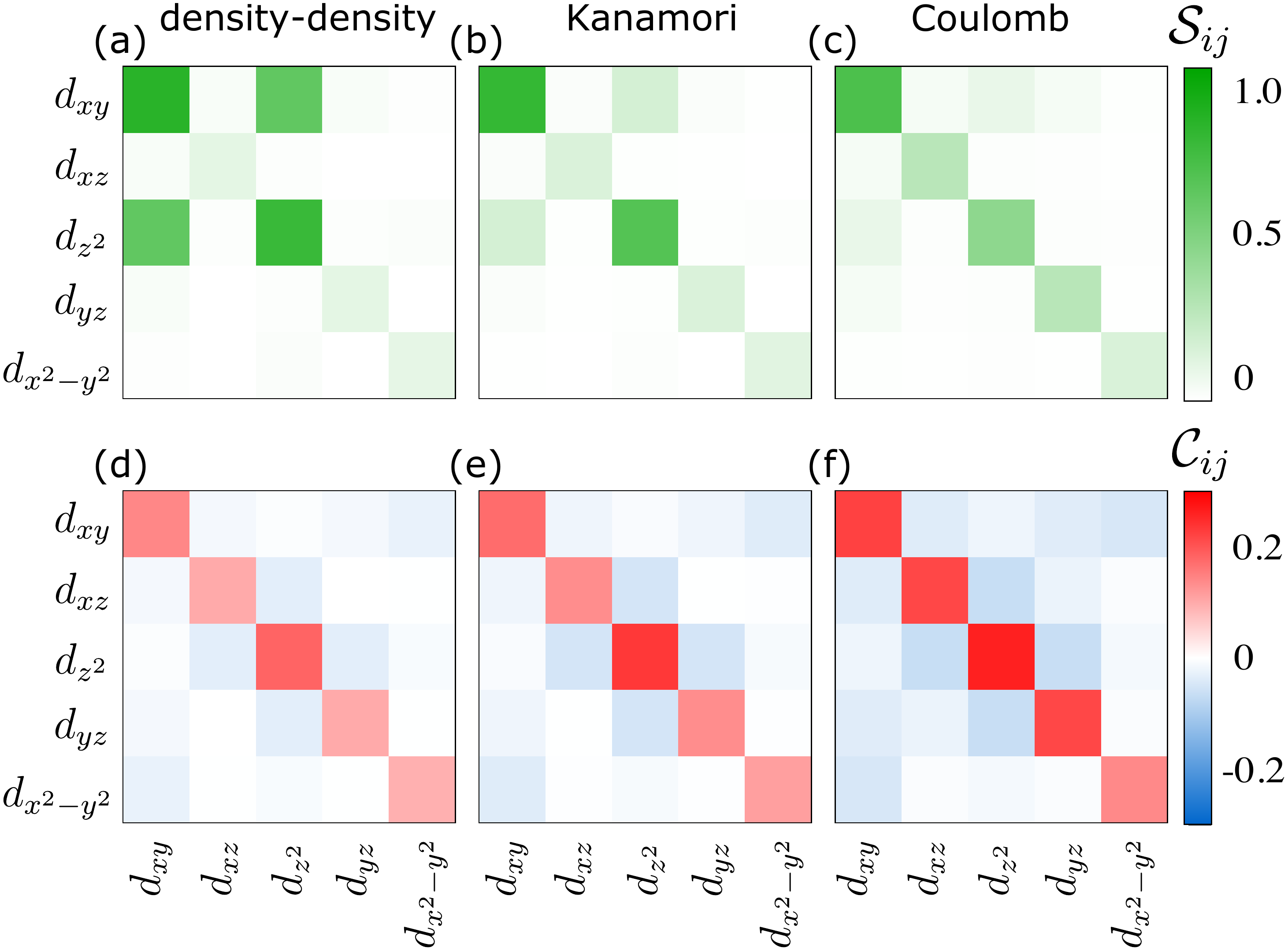}
\caption{Correlation functions describing orbital charge (${\cal C}_{ij}$) and spin (${\cal S}_{ij}$) fluctuations within the Co $3d$ shell, calculated at $T \approx 33$~K. } 
\label{fig:charge-spin_matrix}
\end{figure}

The considerations above can also be better understood
by explicitly calculating the spin and charge fluctuations, defined as 
\begin{eqnarray}
{\cal S}_{ij} & = & \langle \hat{\sigma}_{i} \hat{\sigma}_{j} \rangle
                          - \langle \hat{\sigma}_i \rangle \langle \hat{\sigma}_j \rangle, \\
{\cal C}_{ij} & =&  \langle \hat{n}_{i} \hat{n}_{j} \rangle
                          - \langle \hat{n}_i \rangle \langle \hat{n}_j \rangle,
\end{eqnarray} 
where we introduced the operators  
$\hat{n}_i  = \hat{n}_{i\uparrow} + \hat{n}_{i\downarrow}$ and 
$\hat{\sigma}_i = \hat{n}_{i\uparrow} -  \hat{n}_{i\downarrow}$ 
(and $\langle \hat{\sigma}_{i} \rangle = 0$ in the paramagnetic state).   
In Fig.~\ref{fig:charge-spin_matrix} we show a matrix heatmap 
for each of the correlators above, 
for data representative of the low-temperature regime (at $T\approx33~K)$. 
As usual, we discuss the behavior of spin and charge fluctuations 
upon increasing the complexity of the Coulomb tensor. 
The data support the scenario of the destabilization of the high-spin state 
in the $(d_{xy}, d_{z^2})$ subspace as 
both their spin moments (proportional to the elements $S_{ii}$) 
and their inter-orbital correlator $S_{i \neq j}$ are suppressed, 
while the spin moments of the orbitals in the rest of the $3d$ shell increase, 
as a consequence of the charge redistribution. 
At the same time, we observe a significant enhancement of charge fluctuations, 
in both the inter- and the intra-orbital components  
(in absolute value, as ${\cal C}_{i\neq j} \!<\! 0$). 
The orbital spin polarization is responsible of the orbital decoupling 
in the regime dominated by the Hund exchange~\cite{demediciPRB83,fanfarilloPRB92} 
while the enhancement of the charge fluctuations is the hallmark of increased metallicity 
in the (Kanamori and) full Coulomb parametrization(s),  
as also previously reported in model studies 
of multi-orbital impurity problems~\cite{Huang2014}. 
The high-spin state is weakened 
already by the spin-flip term in the Kanamori Hamiltonian, 
but the two-body mixing terms, involving combinations 
of three (e.g., $U_{ijjk}$) or even four ($U_{ijkl}$) different orbital indices, 
which are included within the full Coulomb parametrization,  
are highly effective in reducing the ``orbital rigidity'' 
and eventually yield a solution which is well described 
neither by a single $S\!=\!1$ Kondo effect~\cite{NCPRL103} 
nor by two independently screened $S\!=\!1/2$ spins~\cite{jaco15,baru15}.
A thorough discussion of these terms 
and their relation with the Hund's coupling is provided in Appendix~\ref{app:tensor}. 

Interestingly, the temperature dependence of both 
spin and charge fluctuations within the Co $3d$ shell 
is negligible with respect to the changes observed 
between different interaction schemes, 
so that the above picture is valid in the whole range $300-30$~K,  
and probably still holds below that. 


\subsection{Spectral signatures of the Kondo effect}\label{sec:spectra}

\begin{figure}[t]
\centering
\includegraphics[width=0.5\textwidth]{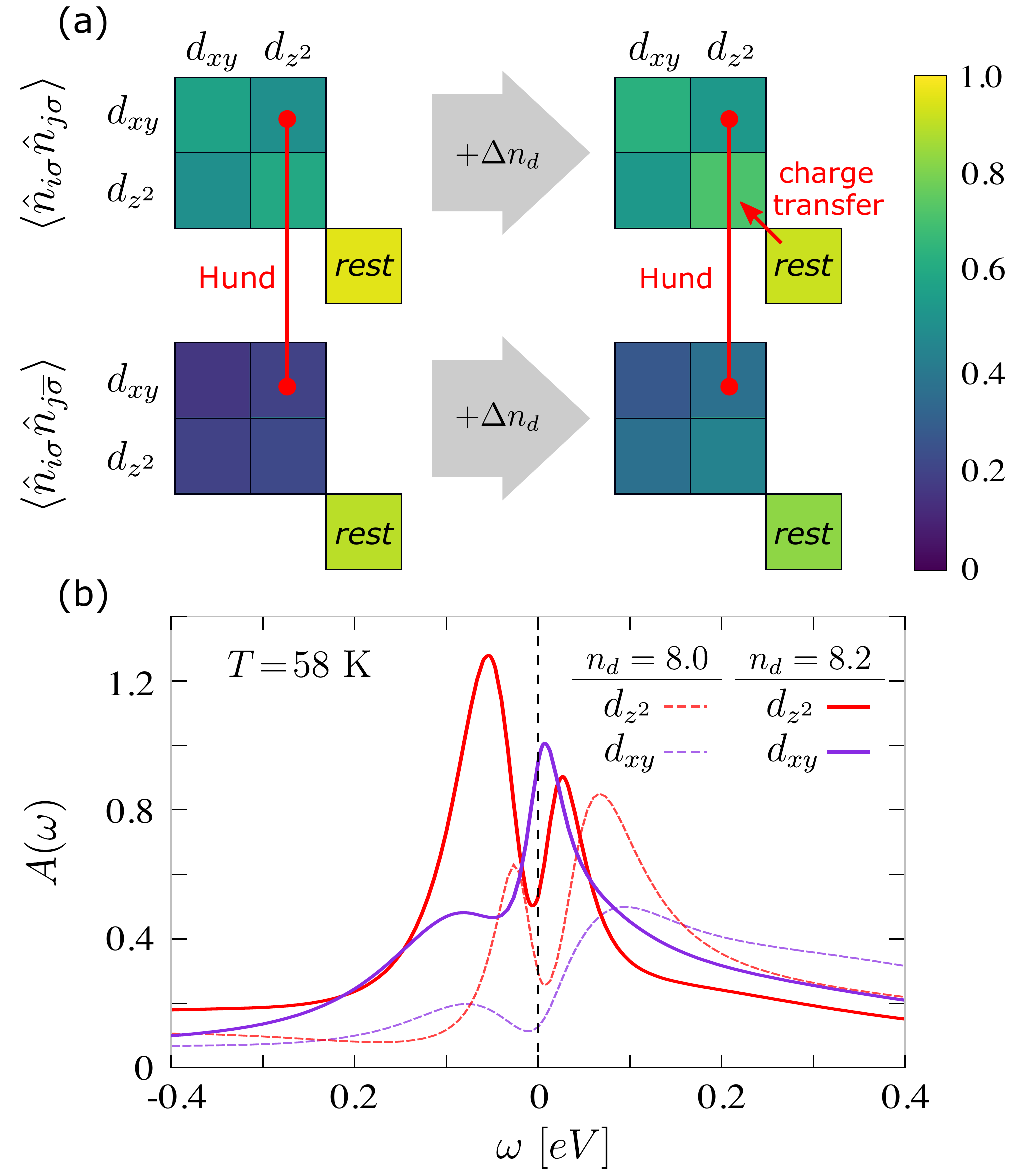}
\caption{(a) Generalized double occupations 
within the density-density approximation. 
The $2 \times 2$ block represents the $(d_{xy}, d_{z^2})$ subspace 
while the extra block, labelled ``rest", denotes 
the average over the diagonal elements 
for the $(d_{yz}, d_{xz}, d_{x^2-y^2})$ subspace. 
Moving away from integer filling $n_d=8$ results in a net charge transfer 
(in addition to the extra $\Delta n_d=0.2$ electrons) 
to the $(d_{xy}, d_{z^2})$ subspace from the rest of the multiplet. 
The overall effect is a weakening of the tendency 
towards the $S=1$ high-spin state. 
(b) Spectral function $A(\omega)$ 
of the $d_{xy}$ and $d_{z^2}$ orbitals within the density-density approximation. 
The development of low-energy resonances away from integer filling 
is compatible with an enhancement of the Kondo scale. }
\label{fig:awdd_charging}
\end{figure}

\begin{figure}[t]
\centering
\includegraphics[width=0.5\textwidth]{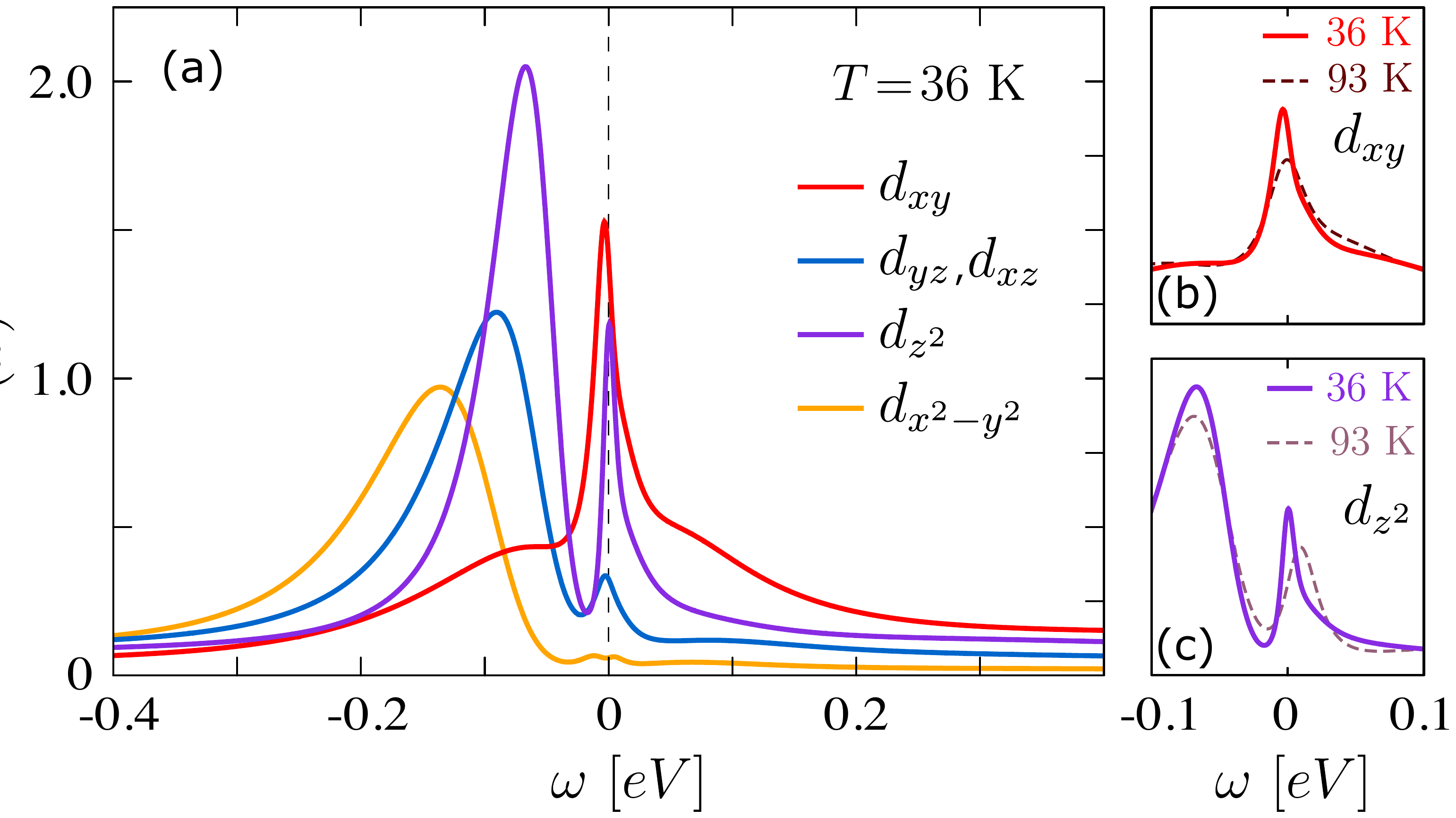}
\caption{[Main panel] Orbital-resolved spectral function 
of the Co $3d$ shell at integer filling, i.e., $n_d=8$ electrons 
in the full Coulomb parametrization. 
The $d_{yz}$ and $d_{xz}$ resonances suggest that  
those orbitals may be relevant to the Kondo effect. 
[Side panels] Temperature evolution of the $d_{xy}$ and $d_{z^2}$ low-energy resonances. }
\label{fig:awc}
\end{figure}

Useful insight can also be obtained by looking at 
the orbital-resolved spectral function of the Co $3d$ shell. 
While one may estimate $T_K$ 
(or at least an \emph{apparent} $T_K$ at $T \neq 0$) 
from spectral features 
such as the width of the resonance~\cite{jaco15,baru15}, 
we will refrain from doing so. 
Since our spectral functions are obtained 
with a numerical analytic continuation procedure 
(maximum entropy method), 
we only take them as \emph{qualitative} indications 
of the redistribution of the spectral weight. 

First, we consider results obtained within the density-density approximation, 
which are shown in Fig.~\ref{fig:awdd_charging}. 
For a Co $3d$ shell occupation of $n_d=8.0$ and at $T = 58$~K, 
neither the $d_{xy}$ nor the $d_{z^2}$ orbital displays 
a resonant feature close to the Fermi level, 
in agreement with the lack of Kondo screening. 
By adjusting to $n_d=8.2$, corresponding to a charge transfer 
from the Cu surface to the Co adatom, prominent resonances 
appear in the spectral functions of both orbitals. 
The analysis of the charge redistribution 
within the $3d$ shell (upper panels of Fig.~\ref{fig:awdd_charging}) 
shows that, upon adding the extra $\Delta n_d=0.2$ electrons,  
the occupation of the Kondo-active orbitals increases as
$(0.57,0.60) \rightarrow (0.63, 0.71)$. 
However, part of the charge accumulating in the subspace comes 
from the rest of the Co $3d$ shell.  
This is indicated by the red arrow in Fig.~\ref{fig:awdd_charging}, 
where the extra block denotes the average occupation 
of the $(d_{yz}, d_{xz}, d_{x^2-y2})$ subspace. 
Such a charge redistribution is detrimental 
to the stabilization of the high-spin state, 
which one realizes by comparing the inter-orbital double occupations 
$\langle \hat{n}_{i\sigma} \hat{n}_{j\sigma}\rangle$ and 
$\langle \hat{n}_{i\sigma} \hat{n}_{j\overline{\sigma}}\rangle$ 
(connected by red dots and a line 
in the upper panels of Fig.~\ref{fig:awdd_charging}).  
The effect of charging is qualitatively analogous to, 
yet not as strong as, what we observed by comparing 
the density-density and full Coulomb parametrizations 
at $n_d=8.0$ in Fig.~\ref{fig:nisnjs}. 
Hence, both scenarios are compatible 
with an enhancement of the Kondo scale.

A clear signature of the Kondo effect is indeed observed 
already at integer filling in the full Coulomb spectral function, 
which is shown in Fig.~\ref{fig:awc} for different temperatures. 
A clear resonance close to the Fermi level 
is observed for both the $d_{xy}$ and the $d_{z^2}$ orbitals. 
The resonance is already present at $T \approx 100$~K, 
but it gets progressively closer to the Fermi level 
and its width decreases as the temperature is lowered 
(see side panels of Fig.~\ref{fig:awc}). 
Interestingly, within this interaction scheme, 
a low-energy resonance develops also in the $(d_{xz}, d_{yz})$ doublet. 
This feature is almost completely absent 
within the density-density approximation, 
and it can be regarded as a further indication 
that a purely $(d_{xy}, d_{z^2})$ description of the Kondo effect 
is not adequate, when accounting for a realistic Coulomb interaction 
in Co/Cu(001). 
Similar resonances are also evident in the spectral functions  
obtained away from integer filling (not shown), 
where the role of the other three orbitals is possibly enhanced.

\section{Discussion and conclusions}\label{sec:discussion}
In this work we investigate the Kondo screening properties 
of Co/Cu(001) in its full realistic complexity. 
We solve an AIM for the whole Co $3d$ shell 
and we focus on the role of the parametrization 
of the Coulomb tensor for the Kondo effect. 
It is important to compare our findings to previous studies 
in the literature, in order to highlight 
both the differences and the similarities. 

Previous theoretical analyses were restricted 
to a two-orbital model for the Kondo-active orbitals~\cite{baru15}, 
with approximate interaction schemes~\cite{baru15,jaco15} 
or impurity solvers~\cite{jaco15}. 
The most direct comparison can be done with the results reported 
by Jacob~\cite{jaco15}, 
obtained with similar interaction parameters as ours, 
derived from first-principles within the constrained random-phase approximation. 
There, many-body effects are taken into account 
at the level of the one-crossing approximation (OCA), 
in contrast to our numerically exact CT-QMC. 
The OCA calculation takes into account all density-density terms 
as well as the spin-flip contributions. 
It may therefore be regarded as an intermediate parametrization 
between density-density and Kanamori, 
albeit restricted to one-crossing diagrams. 
There, a Kondo feature for the $d_{z^2}$ orbital at $T \sim 10$~K for Co/Cu(001), 
is reported, with a Kondo temperature $T_K \approx 90$~K, 
estimated from the width of the Kondo resonance in the spectral function. 
The lack of a similar feature for the other 
Kondo-active orbital ($d_{xy}$ in the notation of this work) 
was suggested as evidence of an underscreened Kondo effect. 
Whether the Co magnetic moment is completely screened 
at lower temperature, with the onset of a Fermi liquid state 
and the realization of a two-stage Kondo effect, 
was not investigated, and it remains debatable. 

On the basis of our CT-QMC results we can delineate 
a quite different situation, whose physical explanation 
can be unveiled thanks to our comparative analysis 
of the various Coulomb tensor parametrizations. 
Within the density-density approximation the overall $T_{K}$ 
is much smaller than the lowest temperature of our calculation,  
and the Nevidomskyy-Coleman scenario for a spin $S=1$ Kondo is fully realized. 
We progressively include additional exchange interactions 
within the Co $3d$ multiplet in the Kanamori 
and eventually all of them in the full Coulomb parametrizations. 
Due to the associated charge redistribution, 
spin fluctuations are partially quenched, 
whereas charge fluctuations increase, 
together with the orbital entanglement. 
Two effects consequently emerge. 
The $d_{xy}$ and $d_{z^2}$ orbitals start to thrive 
on Kondo screening, especially with the full Coulomb interaction, 
while the remaining three $d$-orbitals substantially increase 
their active contribution to the local moment. 
The latter is transparently observed by comparing 
the charge distribution within the Co $3d$ multiplet 
in Fig.~\ref{fig:nisnjs} and the contribution 
of the $(d_{xy}, d_{z^2})$ subspace to the local moment 
in Fig.~\ref{fig:Seff} (by moving from the left to the right panels). 
The relevant role of the whole Co $3d$ shell 
within the full Coulomb parametrization of the interaction 
has also been suggested in the past~\cite{gorelovPRB80,surer11}. 
However, the temperature regime previously investigated 
is hardly relevant for extracting useful information 
about the Kondo screening. 

The outcome of the present study therefore 
changes the conventional interpretation of the Kondo effect 
in the prototypical Co-adatom systems, 
once a realistic interaction tensor is properly taken into account. 
Since the whole $3d$ shell is involved in the Kondo screening, 
one neither has a Nevidomskyy-Coleman scenario 
with the screening of a $S=1$ spin at low temperatures, 
nor two independent $S=1/2$ spin Kondo replicas 
in the $(d_{xy}, d_{z^2})$ subspace. 
The most appropriate way of describing the Kondo effect 
in Co adatoms on a Cu(001) surface is, as a matter of fact, 
a multi-orbital entangled correlated state. 
While two of the five $3d$ orbitals have the largest magnetic moment 
and a favorable hybridization to the substrate 
in order to display clear Kondo peaks, 
they are not decoupled enough from the other orbitals 
to allow for an effective two-orbital description of the Co $3d$ shell. 

Finally, we note that some details of the calculations 
may differ from other results in the literature. 
For instance, Jacob~\cite{jaco15} 
and Baruselli~\textit{et al.}~\cite{baru15} 
consider STM geometries, 
where the STM tip also consists of a Cu pyramid 
grown in the (001) direction -- or the (111) direction 
when considering Co/Cu(111). 
In some cases~\cite{jaco15,pickPRB68,vitaliPRL101}, 
besides the Co-Cu adsorption distance, also the atomic positions 
of some Cu atoms of the surface layer are relaxed. 
Despite these effects possibly being important, 
we are confident that the differences observed 
within the different parametrization of the Coulomb interaction 
influence the Kondo screening in a more fundamental way 
than the details of the DFT calculations. 

To conclude, we revisited the prototypical Co/Cu(001) 
Kondo problem under a new light. 
We established how the parametrization of the Coulomb tensor 
affects the screening of the impurity magnetic moment, 
and we highlight the active role of the whole Co $3d$ shell 
in the Kondo effect. 
Our analysis is likely relevant and can be extended
to other Kondo systems with transition metal adatoms.

\acknowledgments
We thank A.~Amaricci, M.~Capone, L.~Fanfarillo, L.~ de'~Medici, 
A.~Nevidomskyy, M.~Sch\"{u}ler, and T.~O.~Wehling 
for insightful discussions. 
We are also grateful to D.~Jacob for critical reading of the manuscript. 

M. P. B. and C. H. acknowledge the German Research Foundation (DFG) for 
funding via project HE 5675/6-1, the high-performance-computing team 
of the Regionales Rechenzentrum at Universit\"at Hamburg and
the North-German Supercomputing Alliance (HLRN) for computational resources. 
A. K. and G. S. are supported by DFG-SFB 1170 Tocotronics, 
and further acknowledge financial support from the DFG 
through the W\"urzburg-Dresden Cluster of Excellence 
on Complexity and Topology in Quantum Matter ct.qmat 
(EXC 2147, project-id 390858490).
We gratefully acknowledge the Gauss Centre for Supercomputing 
e.V. (www.gauss- centre.eu) for funding this project by providing computing time 
on the GCS Supercomputer SuperMUC at Leibniz Supercomputing Centre (www.lrz.de).
This research was supported in part by the National Science Foundation 
under Grant No. NSF PHY-1748958. 
A.V. acknowledges financial support from the Austrian Science Fund (FWF) 
through the Erwin Schr\"{o}dinger fellowship J3890-N36, 
through project 'LinReTraCe' P~30213, and project P~31631.

\appendix

\section{Details and physical implications of the $k$-mesh convergence 
of the Cu(001) surface} \label{app:kmesh}

\begin{figure}[b]
\centering
\includegraphics[width=0.5\textwidth]{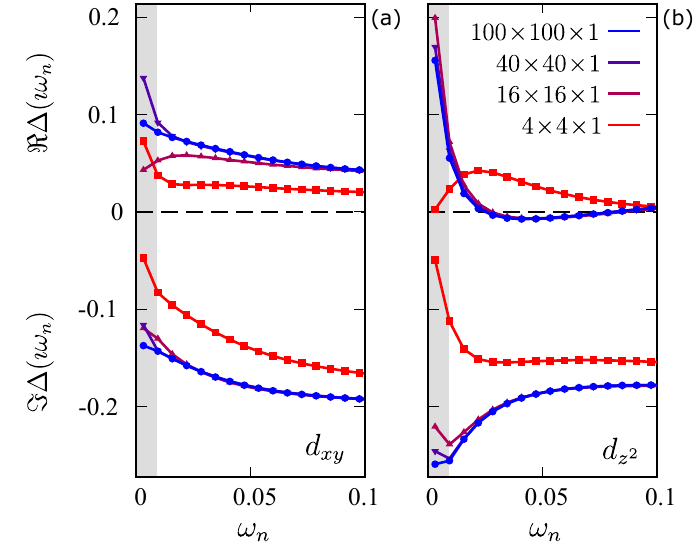}
\caption{Real and imaginary parts of the hybridization function 
of the Co $d_{xy}$ (a) and $d_{z^2}$ (b) orbitals
in the Matsubara representation: $\omega_n=(2n+1)\pi/\beta$, 
with an infra-red cutoff $\propto \beta\!=\!1000$~eV$^{-1}$. 
When evaluated on different $k$-mesh sampling of the Brillouin zone, 
$\Delta(\imath\omega_n)$ shows a slow convergence with the mesh size, 
in particular for the $d_{z^2}$ orbital. }
\label{fig:hyb_k-mesh}
\end{figure}

Throughout our study, we realized that the physical picture 
of Co/Cu(001) delicately depends on the size of the $k$-mesh 
of the Brillouin zone. 
In Fig.~\ref{fig:hyb_k-mesh} we show the orbital-resolved 
hybridization function $\Delta(\imath\omega_n)$ 
describing the embedding of the Co adatom on the Cu surface.  
We compare the results obtained for different $k$-meshes 
(all centered around the $\Gamma$ point). 
We find that $\Delta(\imath\omega_n)$ displays a slow convergence 
with the size of the $k$-mesh, in particular for the $d_{z^2}$ orbital. 
For the sparsest mesh considered, i.e., 
$4 \times 4 \times 1$ $k$-points, 
$\Delta(\imath\omega_n)$ at low frequencies 
displays a \emph{qualitatively} different behavior 
for the $d_{z^2}$ orbital when compared to more accurate meshes 
(with up to $100 \times 100 \times 1$ $k$-points), 
while for the $d_{xy}$ orbital the differences are mainly quantitative. 
Since the differences are observed at relatively low energy scales, 
it is possible that this effect may be overlooked in calculations 
with a low energy resolution, or with a large smearing parameter $\eta$ 
in the hybridization function $\Delta(\omega+\imath\eta)$. 
Differences between the $40 \times 40 \times 1$ and 
$100 \times 100 \times 1$ $k$-points meshes 
can be observed on energy scales $< 0.01$~eV, which corresponds approximatively 
to the lowest Matsubara frequency for the lowest temperatures of our QMC calulations, 
$\beta=350$~eV$^{-1}$ (shaded area in Fig.~\ref{fig:hyb_k-mesh}).

\begin{figure}[t]
\centering
\includegraphics[width=0.5\textwidth]{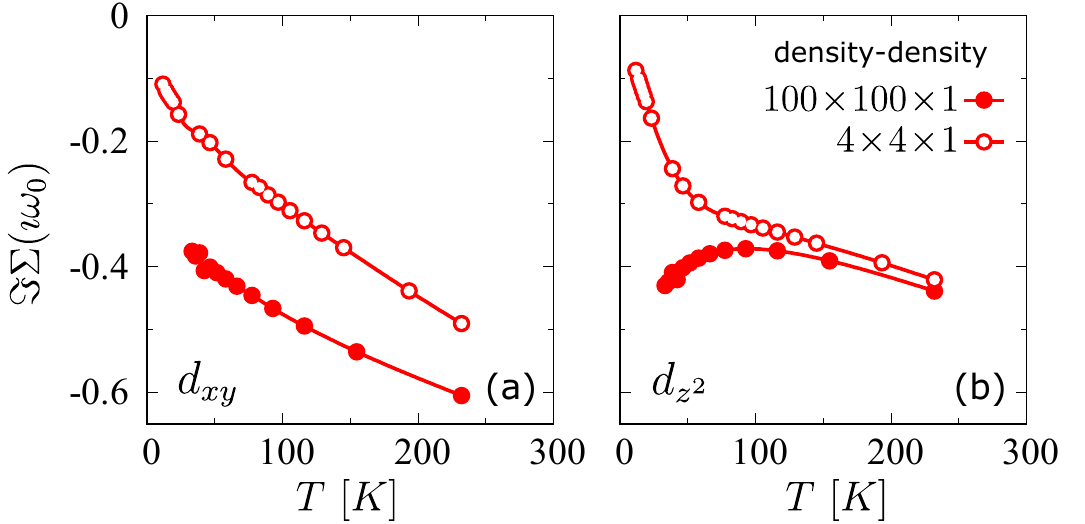}
\caption{Temperature evolution of $\Im \Sigma(\imath\omega_0)$ 
for the Co $d_{xy}$ and $d_{z^2}$ orbitals  within the density-density approximation. 
A sparse $4 \times 4 \times 1$ $k$-mesh suggests the onset 
of a Fermi-liquid regime at relatively high temperature, 
whereas strongly non-Fermi liquid features emerge for denser meshes. }
\label{fig:FMR_mesh}
\end{figure}

However, here we show that this seemingly technical detail 
can have drastic consequences on the physical description of the system. 
For instance, we can consider the temperature evolution 
of the lowest Matsubara frequency $\omega_0$ 
of the electronic self-energy, 
which in a Fermi liquid should scale as 
$\Im \Sigma(\imath\omega_0) \propto T$ (see e.g., Refs.~\cite{Chubukov2012,amaricciPRB85}). 
In Fig.~\ref{fig:FMR_mesh} we show $\Im \Sigma(\imath\omega_0)$ 
for both the $d_{xy}$ and $d_{z^2}$ orbitals 
within the density-density approximation. 
The Fermi liquid scaling seems to be recovered 
at low-enough temperatures 
for the sparsest $4 \times 4 \times 1$ $k$-mesh, for both orbitals. 
For denser $k$-meshes, and in particular 
for the $100 \times 100 \times 1$ one, 
the self-energy of the $d_{z^2}$ orbital 
displays a clear non-Fermi liquid behavior,  
which we follow down to $T \approx 33$~K.

\begin{figure}[t]
\centering
\includegraphics[width=0.5\textwidth]{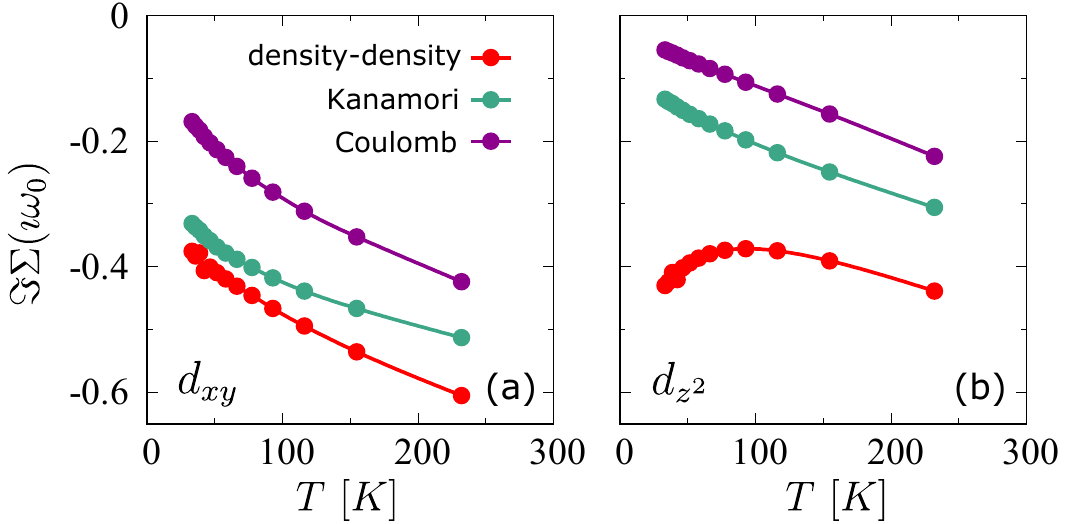}
\caption{Temperature evolution of $\Im \Sigma(\imath\omega_0)$ 
for the Co $d_{xy}$ and $d_{z^2}$ orbitals  
and different parametrization schemes of the Coulomb interaction 
for the $100 \times 100 \times 1$ $k$-mesh. }
\label{fig:FMR}
\end{figure}

Finally, in Fig.~\ref{fig:FMR} we compare $\Im \Sigma(\imath\omega_0)$ 
obtained within all parametrizations of the Coulomb tensor. 
We note that only the full Coulomb case seems to be compatible 
with a linear behavior, although this feature alone is not enough 
to confirm the onset of a Fermi liquid state at low temperatures.  
This observation is important because in the literature, 
calculations for Co/Cu(001) performed 
without including the full Coulomb tensors~\cite{baru15,jaco15} 
indicated the Kondo screening to be the most effective 
for the $d_{z^2}$ orbital. 
Our calculations show that, with an accurate-enough description 
of the hybridization between the Co adatom and the Cu surface, 
and at low-enough temperatures, 
the density-density approximation does not confirm this picture. 
It is instead necessary to take into account more realistic forms 
of the Coulomb interaction to obtain estimates of the Kondo scale 
comparable with the experimental observations. 
 
\section{Coulomb tensor} \label{app:tensor}
The full Coulomb interaction Hamiltonian for the impurity model is given by 
\begin{equation}\label{eq:coulomb}
\hat{H}_{\mathrm{C}} = \frac{1}{2} \sum_{mm'm''m'''} \sum_{\sigma\sigma'} 
                         U_{ijkl} 
                        \hat{d}^{\dag}_{m\sigma} 
                        \hat{d}^{\dag}_{m'\sigma'}
                        \hat{d}^{\phantom{\dag}}_{m'''\sigma'}
                        \hat{d}^{\phantom{\dag}}_{m''\sigma} ,
\end{equation}
where $U_{mm'm''m'''}$ is the Coulomb tensor, labelled by the 
orbital momentum quantum number $m=-\ell, ..., 0, ..., \ell$  
(in this case, $\ell=2$ for the Co $3d$ shell).  
For a spherically symmetric atom, the Coulomb tensor 
can be expressed as follows~\cite{slater1960, Slater1929}
\begin{equation}
 U_{mm'm''m'''} = \sum_{k=0}^{2\ell} a_{k}(mm';m''m''')F^k, 
\end{equation}
where the coefficients $a_k$ and the Slater parameters $F^k$ are given by integrals 
of spherical harmonics and radial part of the wave function, respectively.  
Their expressions are well known and can be found, 
e.g., in Refs.~\cite{slater1960, Slater1929, karolakPhD}.
In the basis of the spherical harmonics, the Coulomb tensor includes 
all two-, three-, and four-index interaction terms 
which separately fulfill the conservation of both spin and angular momentum~\cite{slater1960}. 
The first condition is encoded in the choice 
of the spin indices of Hamiltonian~(\ref{eq:coulomb}), 
while the latter reads $m+m'=m''+m'''$.  

For actual calculations it is convenient to rotate the Coulomb tensor 
$U_{mm'm''m'''} \rightarrow U_{ijkl}$ from spherical to cubic harmonics. 
The corresponding transformation ($m>0$) for the basis functions is given by
\begin{equation}
 \begin{split}
  K_{\ell}^{m} &= \frac{1}{\sqrt{2}}\Big[ (-1)^m Y_{\ell}^{m} + Y_{\ell}^{-m} \Big] \\
  K_{\ell}^{0}  &= Y^{\ell}_{0} \\
  K_{\ell}^{-m} &= \frac{1}{\imath\sqrt{2}}\Big[ (-1)^m Y_{\ell}^{m} - Y_{\ell}^{-m} \Big]. 
 \end{split}
\end{equation}
In the case of the Co $3d$ shell we label the cubic harmonics as 
$(K_{2}^{-2}, K_{2}^{-1}, K_{2}^{0}, K_{2}^{1}, K_{2}^{2}) = 
         (d_{xy}, d_{xz}, d_{z^2}, d_{yz}, d_{x^2-y^2})$. 
A data file containing the full Coulomb tensor in this basis, 
which we used in all of our numerical calculations, is also provided as Supplementary Material.

\subsection{One- and two-index interaction terms}
The terms of the Coulomb tensor $U_{ijkl}$ which contain only two different indices
can be classified as follows. 
The density-density terms, which include the intra-orbital interaction $U_{iiii}$, 
(existing only for $\sigma \neq \sigma'$ due to the Pauli exclusion principle), 
and the inter-orbital interaction $U_{ijij}=(U_{iiii}+U_{jjjj})/2-2U_{ijji}$, 
where $U_{ijji}$ represents the Hund's exchange coupling 
for parallel spin configurations~\cite{jaco15}. 
Other additional exchange terms, for opposite spin configurations, 
account for spin-flip and pair hopping processes, 
with values $U_{ijji}$ and $U_{iijj}$, which are the same as 
the density-density exchange in the cubic harmonics basis. 

It is possible to take all previous terms into account in a relatively compact form, 
giving rise to the Kanamori interaction Hamiltonian 
[see also Eqs.~(\ref{eq:density-density}, \ref{eq:kanamori})] 
\begin{equation}\label{eq:kanamori_full}
 \begin{split}
  \hat{H}_{\mathrm{K}} = 
   & \sum_{i} 
         U_{ii}
         \hat{n}_{i\uparrow}\hat{n}_{i\downarrow} 
  + \sum_{i \neq j} \sum_{\sigma\sigma'} 
        (U_{ij}-J_{ij}\delta_{\sigma\sigma'}) 
         \hat{n}_{i\sigma} \hat{n}_{j\sigma'} \\
  + & \sum_{i\neq j} J_{ij}
        \big(
             \hat{d}^{\dag}_{i\uparrow}
             \hat{d}^{\dag}_{j\downarrow}
             \hat{d}^{\phantom{\dag}}_{i\downarrow}
             \hat{d}^{\phantom{\dag}}_{j\uparrow} 
           - \hat{d}^{\dag}_{i\uparrow}
             \hat{d}^{\dag}_{i\downarrow}
             \hat{d}^{\phantom{\dag}}_{j\uparrow}
             \hat{d}^{\phantom{\dag}}_{j\downarrow}
        \big) ,
 \end{split}
\end{equation}
which is defined in terms of the two-index interactions 
$U_{ii}=U_{iiii}$, $U_{ij}=U_{ijij}$, and $J_{ij} = U_{ijji}$. 
One can reduce to only two parameters, $U$ and $J$, by expressing the interaction 
in terms of the Slater integrals $F^0$, $F^2$, and $F^4$, as
\begin{eqnarray}
U_0    & = & F^0 + \frac{8}{7} \frac{1}{14}(F^2+F^4) \\ 
J_1 & = & \frac{1}{49}(3F^2 + \frac{20}{9} F^4) \\
J_2 & = & -2 \ \frac{5}{7} \frac{1}{14}(F^2+F^4) + 3 J_1 \\
J_3 & = & 6 \ \frac{5}{7} \frac{1}{14}(F^2+F^4) - 5 J_1 \\
J_4 & = & 4 \ \frac{5}{7} \frac{1}{14}(F^2+F^4)  - 3 J_1,
\end{eqnarray}
and identifying $U=F^0$ and $J=\frac{1}{14}(F^2+F^4)$ 
with an almost constant ratio $F^4/F^2 \approx 0.625$ for $3d$ ions~\cite{schnellPRB68}.  
See also, e.g., Refs.~\cite{hausoelNC8,karolakPhD} for a related discussion. 

\begin{figure}[t]
\centering
\includegraphics[width=0.4\textwidth]{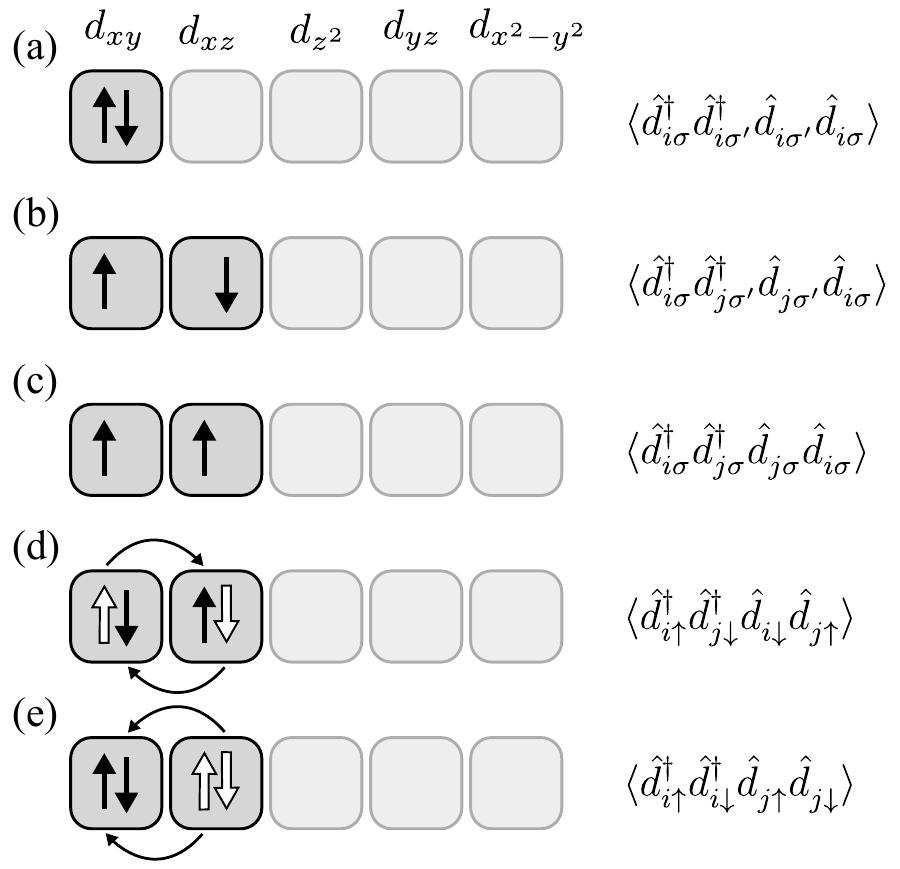}
\caption{Schematic representation of the two-index interaction terms 
(excluding permutations) for the $3d$ shell in the basis of the cubic harmonics.  
The white and black spins denote the initial and final configurations 
connected by the operator, respectively. 
The terms are: 
(a) intra-orbital interaction, (b, c) inter-orbital interactions, including the Hund's exchange 
for parallel spin configurations, (d) spin-flip, (e) pair-hopping. }
\label{fig:umatrix_two-index}
\end{figure}

\noindent 
All two-index terms in the basis of the $3d$ cubic harmonics 
are summarized in Tables~\ref{tab:Uijji} and~\ref{tab:Uijij}, for reference. 
In Fig.~\ref{fig:umatrix_two-index} we show a schematic representation 
of all possible two-index interaction terms (excluding permutations) 
for the $3d$ shell in the basis of the cubic harmonics.

For our spherically-symmetric calculations of the Co/Cu(001) system,   
we set $U=4.5$~eV and $J=0.7$~eV, which characterize the interaction matrix 
with Slater integrals 
$F^0 = 4.0$~eV, $F^2  \simeq 7.75$~eV, and $F^4  \simeq 4.85$~eV,  
and result in the interaction parameters 
$U_0=5.02$~eV, $J_1=0.69$~eV, $J_2=0.80$~eV, $J_3=0.39$~eV, $J_4=0.49$~eV.

\begin{table}[t]
\caption{Intra- and inter-orbital interactions of density-density type 
in the spherically symmetric Coulomb tensor. }
\label{tab:Uijji}
\begin{ruledtabular}
\begin{tabular}{clllll}
 $U_{ijij}$ & $d_{xy}$ & $d_{xz}$ & $d_{z^2}$ & $d_{yz}$ & $d_{x^2-y^2}$ \\ 
\colrule
$d_{xy}$          & $U_0\phantom{-2J_1}$ & $U_0-2J_1$ & $U_0-2J_2$ & $U_0-2J_1$ & $U_0-2J_3$ \\
$d_{xz}$          &        & $U_0\phantom{-2J_1}$           & $U_0-2J_4$ & $U_0-2J_1$ & $U_0-2J_1$ \\
$d_{z^2}$        &        &                  & $U_0\phantom{-2J_1}$           & $U_0-2J_4$ & $U_0-2J_2$ \\
$d_{yz}$          &        &                  &                   & $U_0\phantom{-2J_1}$         & $U_0-2J_1$ \\
$d_{x^2-y^2}$ &        &                   &                  &	           & $U_0\phantom{-2J_1}$           \\
\end{tabular}
\end{ruledtabular}
\end{table}

\begin{table}[t]
\caption{Hund's exchange couplings of density-density and Kanamori type 
in the spherically symmetric Coulomb tensor.}
\label{tab:Uijij}
\begin{ruledtabular}
\begin{tabular}{clllll}
 $U_{ijji}$ & $d_{xy}$ & $d_{xz}$ & $d_{z^2}$ & $d_{yz}$ & $d_{x^2-y^2}$ \\ 
\colrule
$d_{xy}$          & - & $J_1$ & $J_2$ & $J_1$ & $J_3$ \\
$d_{xz}$          &   &   -        & $J_4$ & $J_1$ & $J_1$ \\
$d_{z^2}$        &   &            &  -         & $J_4$ & $J_2$ \\
$d_{yz}$          &   &            &            & -          & $J_1$ \\
$d_{x^2-y^2}$ &   &             &            &	      & -          \\
\end{tabular}
\end{ruledtabular}
\end{table}

\noindent 
In the case of a non-spherical Coulomb tensor, 
some symmetries between the interaction terms are lifted. 
For instance, symmetries between all pairs of
$d_{xy}$, $d_{xz}$, and $d_{yz}$ orbitals, 
or between the $d_{z^2}$ and any of the two planar orbitals ($d_{xy}$ and $d_{x^2-y^2}$).  
Moreover, the intra-orbital term $U_{iiii}$ becomes 
orbital-dependent. 
The non-spherical interaction parameters for Co/Cu(001) have been evaluated 
with the constrained random phase approximation~\cite{jaco15}.

\begin{figure}[t]
\centering
\includegraphics[width=0.4\textwidth]{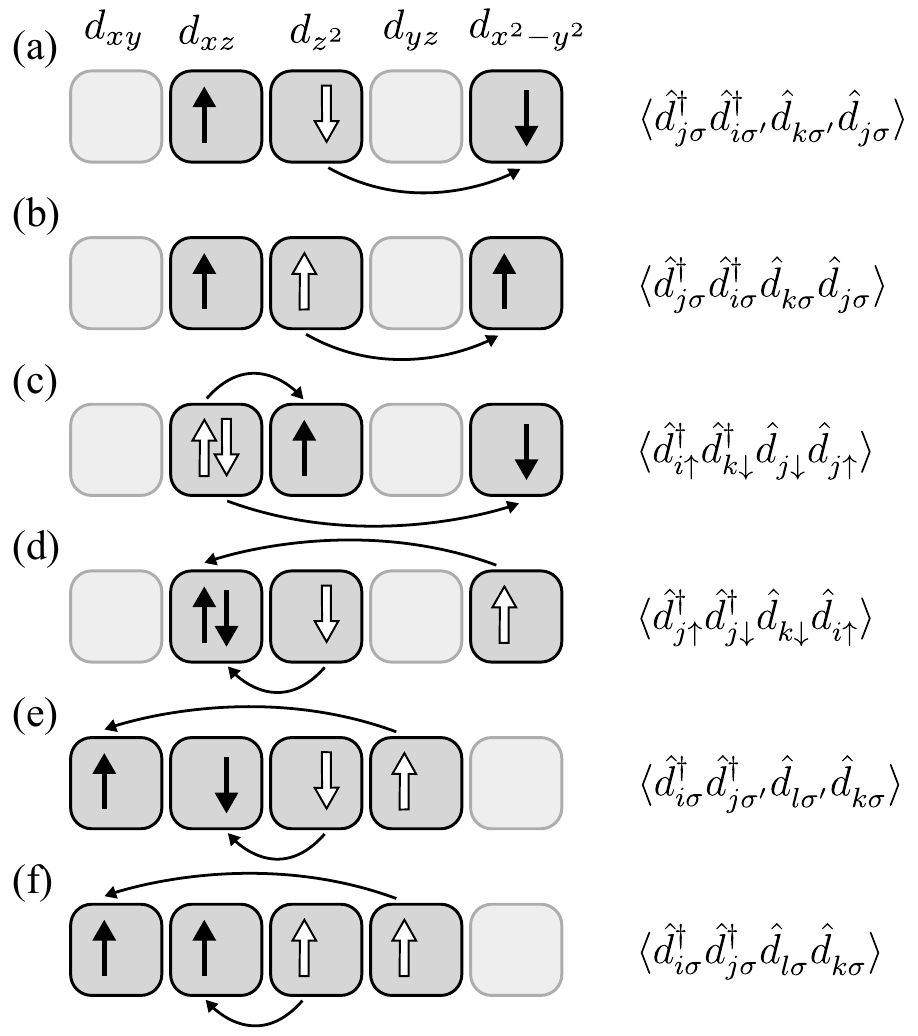}
\caption{Schematic representation of the three- and four-index interaction terms 
(excluding permutations) for the $3d$ shell in the basis of the cubic harmonics.  
The white and black spins denote the initial and final configurations 
connected by the operator, respectively. 
The terms are: 
(a, b) correction to the Hund's exchange due to an inter-orbital effective hopping, 
(c, d) annihilation and creation of an orbital pair, 
(e, f) two-unpaired-electrons hopping. }
\label{fig:umatrix_three-four-index}
\end{figure}

\subsection{Three- and four-index interaction terms}
Even though the Kanamori parametrization restores the rotational invariance 
of the Coulomb tensor (in the spherical approximation) and provides 
an exact parametrization for two- and three-orbital models, 
it does not contain all possible interaction terms allowed for the whole $3d$ shell.

In the basis of the spherical harmonics, there exist three-index terms for the form 
$U_{mmm'm''}$ and $U_{m'm''mm}$ such that $2m = m'+m''$. 
All other terms, e.g., $U_{mm'm''m}$ or $U_{m'mmm''}$ 
can only conserve the angular momentum if $m'=m''$, 
giving rise to two-index terms already included in the Kanamori parametrization.
These interactions can be interpreted in terms of the creation or annihilation of an orbital pair. 
However, once rotated in the cubic harmonics basis, 
besides the pair creation ($U_{jjik}$) or annihilation ($U_{ikjj}$) terms, 
one also obtains terms associated to matrix elements 
$U_{jijk}$ and $U_{jikj}$ ($\sigma=\sigma')$. 
Similarly, the only four-index term allowed 
the pair of annihilation operators can only carry 
angular momentum $m + m' = \pm 1$ or $m + m' = 0$ ($m \neq 0$), 
which is mirrored by the creation operators. 
These interactions resembles hopping of unpaired electrons 
involving four different orbitals 
both in the spherical and in the cubic harmonics basis. 
In Fig.~\ref{fig:umatrix_three-four-index} we show a schematic representation 
of three and four-index interaction terms (excluding permutations) 
for the $3d$ shell in the basis of the cubic harmonics. 

A complete "analytic" parametrization of all possible three- and four-index terms, 
in analogy to the standard one for the two-index terms, is out of the scope of this work. 
However, all the integrals $a_k$ are tabulated~\cite{slater1960}  
and the numerical values of the corresponding terms can be calculated 
in a straightforward way for the $3d$ shell, 
given the Slater integrals $F^0$, $F^2$, and $F^4$. 
For the values chosen here, we obtain three independent parameters 
in the spherical approximation, which we refer to as 
$J_5=0.18$~eV, $J_6=0.35$~eV, $J_7=0.31$~eV. 
These interactions appear with both positive and negative sign 
in the Coulomb tensor, and they are associated to three- and four-index terms, 
so that a complete disentangling of these contributions is a non-trivial task, 
even in the spherical approximation. 
We provide a datafile with all elements of the full Coulomb tensor $U_{ijkl}$, 
in the basis of the cubic harmonics, 
in the Supplementary Material.

\section{Quantum Numbers}
It is worth mentioning that all calculations for the Co/Cu(001) system 
have been performed with the full Coulomb tensor, 
and the interaction Hamiltonian (i.e., density-density, Kanamori, or full Coulomb) 
is selected by requiring the conservation of a specific set of quantum numbers. 

Any spin-independent two-body interaction conserves  
the electron number $\sum_{i\sigma} \hat{n}_{i\sigma}$ 
and the spin projection $\hat{S}_z$. 
The density-density interaction conserves 
the electron number in each spin-orbital $\hat{n}_{i\sigma}$.  
Instead, the Kanamori interaction conserves the quantity 
$\sum_i 2^i (\hat{n}_{i\uparrow}-\hat{n}_{i\downarrow})^2$, 
which represents the pattern of orbital single occupations, 
also known as PS number~\cite{Parragh2012}, 
but relaxes the conservation of the electron number on each orbital,  
regardless of spin, by allowing the spin-flip exchange interaction term 
(see schematics in Fig.~\ref{fig:umatrix_two-index}).  
In general three- and four-index interaction terms included in the full Coulomb tensor 
conserve, e.g., nor the spin-orbital occupation nor the PS number 
(see schematics in Fig.~\ref{fig:umatrix_three-four-index}).

In the CT-QMC calculations for the simpler interaction parametrizations using
\textsc{w2dynamics}~\cite{Wallerberger2018}, the level of simplification 
was specified by requiring the conservation of the
appropriate quantum numbers. This causes the local state
space to be partitioned in such a way that terms of the Coulomb tensor connecting
states with different quantum number values do not enter into the
imaginary time evolution.~\cite{Wallerberger2018}

\bibliographystyle{apsrev}

\end{document}